\documentclass[aps,prx,twocolumn,superscriptaddress]{revtex4-2}

\usepackage{graphicx}

\newcommand{\upperRomannumeral}[1]{\uppercase\expandafter{\romannumeral#1}}

\usepackage{hyperref}
\pdfoutput=1
\usepackage{comment}
\usepackage{multirow}
\usepackage{array}
\usepackage{amsmath}
\usepackage{bm}
\usepackage{amssymb}
\usepackage{textcomp}
\usepackage{stmaryrd} 
\usepackage{ulem}
\usepackage{gensymb}

\usepackage{import}
\usepackage{color}
\usepackage{verbatim}

\usepackage{float}

\usepackage{lipsum}
\usepackage{sistyle} 

\newcommand{\qpar}{q}

\def\XXint#1#2#3{{\setbox0=\hbox{$#1{#2#3}{\int}$}
     \vcenter{\hbox{$#2#3$}}\kern-.5\wd0}}

\begin{document}

\title{Sub-nm range momentum-dependent exciton transfer\\ from a 2D semiconductor to graphene}

\author{Aditi Raman Moghe}
\affiliation{Universit\'e de Strasbourg, CNRS, Institut de Physique et Chimie des Mat\'eriaux de Strasbourg, UMR 7504, F-67000 Strasbourg, France}

\author{Delphine Lagarde}
\affiliation{Universit\'e de Toulouse, INSA-CNRS-UPS, LPCNO, 135 Avenue de Rangueil, 31077, Toulouse, France}

\author{Sotirios Papadopoulos}
\affiliation{Universit\'e de Strasbourg, CNRS, Institut de Physique et Chimie des Mat\'eriaux de Strasbourg, UMR 7504, F-67000 Strasbourg, France}

\author{Etienne Lorchat}
\affiliation{Universit\'e de Strasbourg, CNRS, Institut de Physique et Chimie des Mat\'eriaux de Strasbourg, UMR 7504, F-67000 Strasbourg, France}

\author{Luis E. Parra L\'opez}
\affiliation{Universit\'e de Strasbourg, CNRS, Institut de Physique et Chimie des Mat\'eriaux de Strasbourg, UMR 7504, F-67000 Strasbourg, France}

\author{Lo\"ic Moczko}
\affiliation{Universit\'e de Strasbourg, CNRS, Institut de Physique et Chimie des Mat\'eriaux de Strasbourg, UMR 7504, F-67000 Strasbourg, France}


\author{Kenji Watanabe}
\affiliation{Research Center for Functional Materials, National Institute for Materials Science, 1-1 Namiki, Tsukuba 305-0044, Japan }

\author{Takashi Taniguchi}
\affiliation{ International Center for Materials Nanoarchitectonics, National Institute for Materials Science, 1-1 Namiki, Tsukuba 305-0044, Japan }

\author{Michelangelo Romeo}
\affiliation{Universit\'e de Strasbourg, CNRS, Institut de Physique et Chimie des Mat\'eriaux de Strasbourg, UMR 7504, F-67000 Strasbourg, France}

\author{Maxime Mauguet}
\affiliation{Universit\'e de Toulouse, INSA-CNRS-UPS, LPCNO, 135 Avenue de Rangueil, 31077, Toulouse, France}

\author{Xavier Marie}
\affiliation{Universit\'e de Toulouse, INSA-CNRS-UPS, LPCNO, 135 Avenue de Rangueil, 31077, Toulouse, France}
\affiliation{Institut Universitaire de France, 75231 Paris, France}

\author{Arnaud Gloppe}
\affiliation{Universit\'e de Strasbourg, CNRS, Institut de Physique et Chimie des Mat\'eriaux de Strasbourg, UMR 7504, F-67000 Strasbourg, France}

\author{C\'edric Robert}
\affiliation{Universit\'e de Toulouse, INSA-CNRS-UPS, LPCNO, 135 Avenue de Rangueil, 31077, Toulouse, France}

\author{St\'ephane Berciaud}
\email{stephane.berciaud@ipcms.unistra.fr}
\affiliation{Universit\'e de Strasbourg, CNRS, Institut de Physique et Chimie des Mat\'eriaux de Strasbourg, UMR 7504, F-67000 Strasbourg, France}


\begin{abstract}

Heterostructures made from atomically thin semiconductors (here MoSe$_2$) and graphene are uniquely poised to investigate photoinduced charge and energy transfer in the 2D limit. Here, using picosecond time-resolved photoluminescence spectroscopy at cryogenic temperatures on two types of MoSe$_2$/graphene heterostrutures, we unveil key features of the underlying mechanisms. First, the shortening of the MoSe$_2$ bright exciton lifetime is marginally affected by the number of graphene layers to which MoSe$_2$ is coupled. Second, exciton transfer vanishes when a sub-nm thick spacer of hexagonal boron nitride decouples MoSe$_2$ from graphene. These results indicate that charge tunneling govern bright exciton relaxation in MoSe$_2$/graphene and that longer-range, F\"orster-type energy transfer (FRET) does not affect bright excitons. However, sub-ps FRET to graphene accelerates the relaxation of ``hot'' excitons formed upon optical excitation, leading to photoluminescence quenching factors that exceed expectations based on the shortening of the bright exciton lifetime. Our work has direct implications for energy harvesting and funneling using van der Waals heterostructures.

\end{abstract}

\maketitle


\textbf{Introduction --}
A wealth of low-dimensional physical phenomena and related models can be quantitatively tested using atomically-thin layers of van der Waals materials. These materials enable investigations of truly two-dimensional (2D) physics, without being hindered by issues associated with dangling bonds, buffer layers and surface roughness~\cite{Novoselov2016}. In particular, it is now clear that interfacial coupling may strongly alter the electronic, vibrational and optical response of coupled layers~\cite{Hong2014,Rivera2015,Pospischil2013,Raja2017,Moczko2025}. Interlayer (spatially-indirect) excited states may form on ultrafast timescales upon photoexcitation in heterobilayers of transition dichalcogenides (TMD)~\cite{Hong2014,Rivera2015,Kozawa2016}. Correlated electronic phases emerge in twisted layers of graphene~\cite{Cao2018}, as well as in twisted TMD homo- or hetero-bilayers~\cite{Wilson2021}. Coupled layers also offer a model system to investigate the role of dielectric screening~\cite{Raja2017,Raja2019,Gerber2018,Tebbe2023,Moczko2025} as well as charge and energy transfer in a layered donor-acceptor system separated by a sub-nm van der Waals gap. Recently, ultrafast interlayer charge and/or energy transfer phenomena has been reported using a variety of time-resolved (TR) experimental techniques such as transient absorption spectroscopy~\cite{He2014,Zhou2021}, TR photocurrent spectroscopy~\cite{Massicotte2016}, TR photoluminescence (TRPL) spectroscopy~\cite{Kozawa2016,Froehlicher2018,Lorchat2020}, TR Raman spectroscopy~\cite{Ferrante2022} TR angle-resolved photo-emission spectroscopy (TR-ARPES)~\cite{Aeschlimann2020,Krause2021,Dong2023}. However, fundamental questions remain open regarding the microscopic transfer mechanisms at play, specifically the range of the interactions, the associated timescales and the different behaviors between bright excitons (with near-zero center of mass momentum),  hot excitons (optically inactive, with finite center of mass momentum) and charge carriers.

On the theoretical side, it becomes challenging and often speculative to address quantitatively the charge and energy transfer rates in the limit of sub-nanometer separation between coupled layers, where orbital overlap prevail~\cite{Koppens2011,Selig2019,Dong2023}. It is thus important to scrutinize the specific case of tightly coupled TMD and graphene layers in order to get an accurate estimation of the transfer time as well as clear insights into the transfer mechanism.

In this work, we study the low-temperature excitonic dynamics in two distinct types of MoSe$_2$/graphene heterostructures. First, monolayers of MoSe$_2$ directly coupled to ``straircase-like'' graphene flakes comprising several domains with distinct thicknesses. We observe two effects: i) sizeable PL quenching and ii) reduction of the bright exciton lifetime. Remarkably, the latter is marginally dependent on the number of graphene layers $N$, whereas PL quenching shows an appreciable increase with $N$. Second, investigations of MoSe$_2$ monolayers decoupled from a graphene monolayer by  ultrathin hexagonal boron nitride (hBN) spacers reveals that a two layer-thick hBN spacer suffices to fully decouple MoSe$_2$ from graphene. These results strongly indicate that the dominant transfer mechanism for bright excitons is very short range (sub-nm) and implies charge tunneling. In addition, by comparing our experimental results to an electrodynamic model~\cite{SMnote}, we argue that even if dipole-dipole mediated (F\"orster-type) energy transfer (FRET) has no sizeable impact on bright excitons, this mechanism contributes to PL quenching by accelerating the relaxation of hot excitons through efficient transfer to graphene layers, with a rate that appreciably depends on $N$. Our work sheds new light on charge and energy transfer mechanisms at ultrashort (sub-nm) distances and is a strong impetus to further develop existing theoretical frameworks to address charge and energy transfer in the charge tunneling regime.

~

\textbf{PL of MoSe$_2/N$LG and MoSe$_2/N$L\:hBN/1LG --}
MoSe$_2$ is one of the best documented layered semiconductors. Monolayer MoSe$_2$ (thereafter simply denoted as MoSe$_2$) has a direct bandgap and displays a lowest lying optically active (bright) excitonic state ($\mathrm X^0$)~\cite{Ross2013,Wang2018}. Thus, the excitonic emission characteristics of MoSe$_2$ are relatively simple as compared to other materials such as WS$_2$ and WSe$_2$, which exhibit optically inactive (dark) or weakly active (gray) states below the bright excitonic state~\cite{Robert2017}. 

\begin{figure}[t!]
    \centering
    \includegraphics[width=1\linewidth]{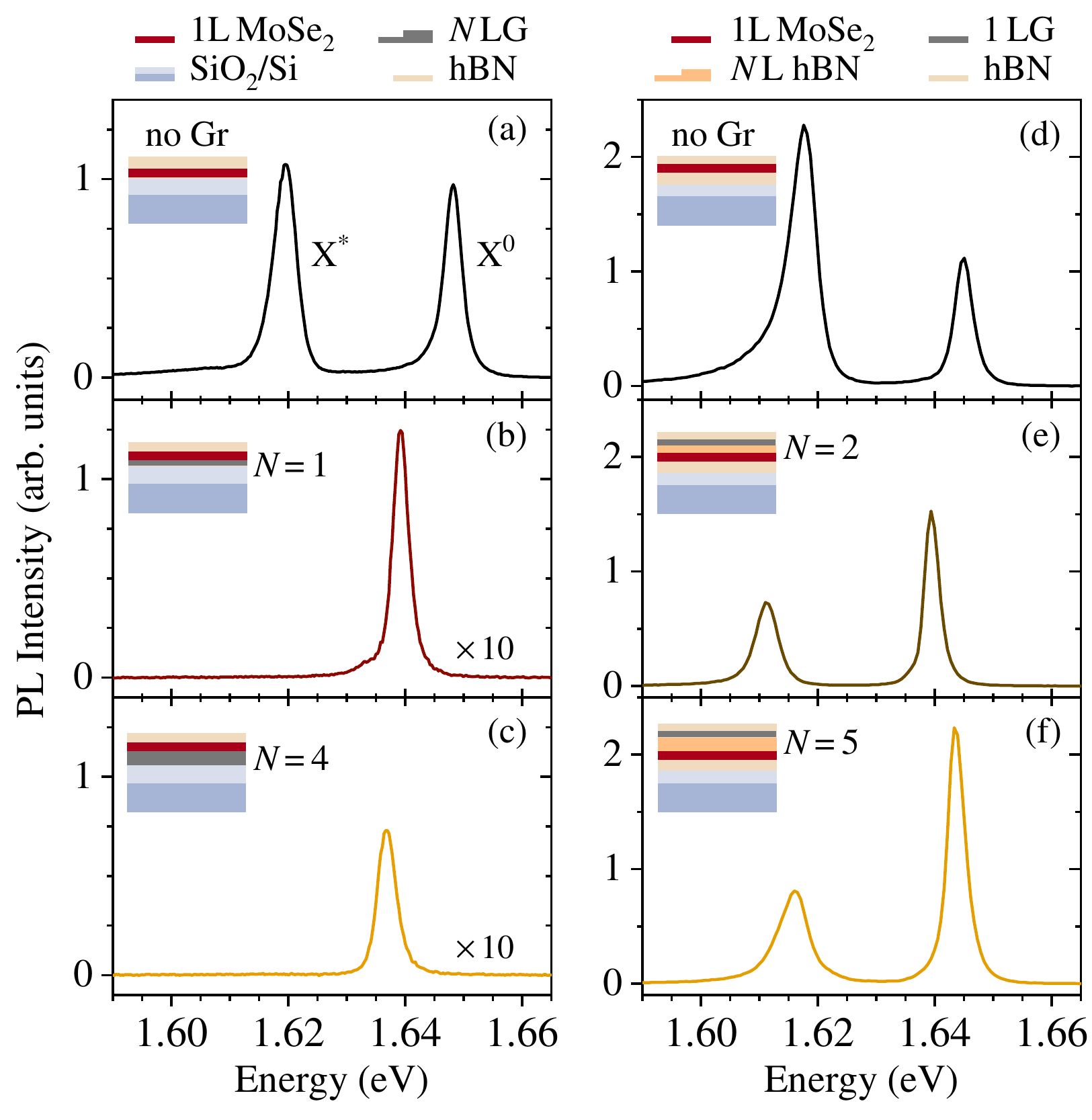}
    \caption{(a-c) Sample 1, featuring (a) an SiO$_2$/MoSe$_2$/hBN reference, (b) 1LG/MoSe$_2$/hBN and (c) 4LG/MoSe$_2$/hBN domains. (d-f) Sample 2, featuring (d) an hBN/MoSe$_2$/hBN reference, (e) hBN/MoSe$_2$/2L\:hBN/1LG and (f) hBN/MoSe$_2$/5L\:hBN/1LG domains. The stacking sequence is sketched in each panel. PL spectra were recorded at cryogenic temperature (16\:K in (a-c) and 4\:K in (d-f)), under continuous wave excitation, in the linear excitation regime. The bright exciton $(\mathrm X^0)$ and trion $(\mathrm X^{\star})$ are indicated in (a).} 
    \label{Fig1}
\end{figure}

Figure~\ref{Fig1} shows low temperature PL spectra of the two types of heterostructures introduced above, namely an hBN-capped MoSe$_2/N$LG heterostructure (Sample 1, Fig.~\ref{Fig1}a-c, with $N=1,\:2,\:3,\:4,\:\mathrm {and}~6$) and an hBN-encapsulated MoSe$_2/N$L\: hBN/1LG heterostructure (Sample 2, Fig.~\ref{Fig1}d-f, with $N=2\: \mathrm {and}\: 5$). All samples were made using standard exfolation and transfer methods~\cite{Novoselov2016,SMnote}. Crucially, and as discussed in details below, we have chosen the  thickness of the SiO$_2$ epilayer and of the bottom hBN layer to control the $\mathrm X^0$ radiative lifetime through cavity effects~\cite{Fang2019,SMnote}. Since we are investigating near-field phenomena that are governed by interfacial coupling, the quality and homogeneity of the van der Waals heterointerfaces in our samples has been thoroughly characterized~\cite{SMnote}.

\begin{figure}[ht!]
    \centering
    \includegraphics[width=\linewidth]{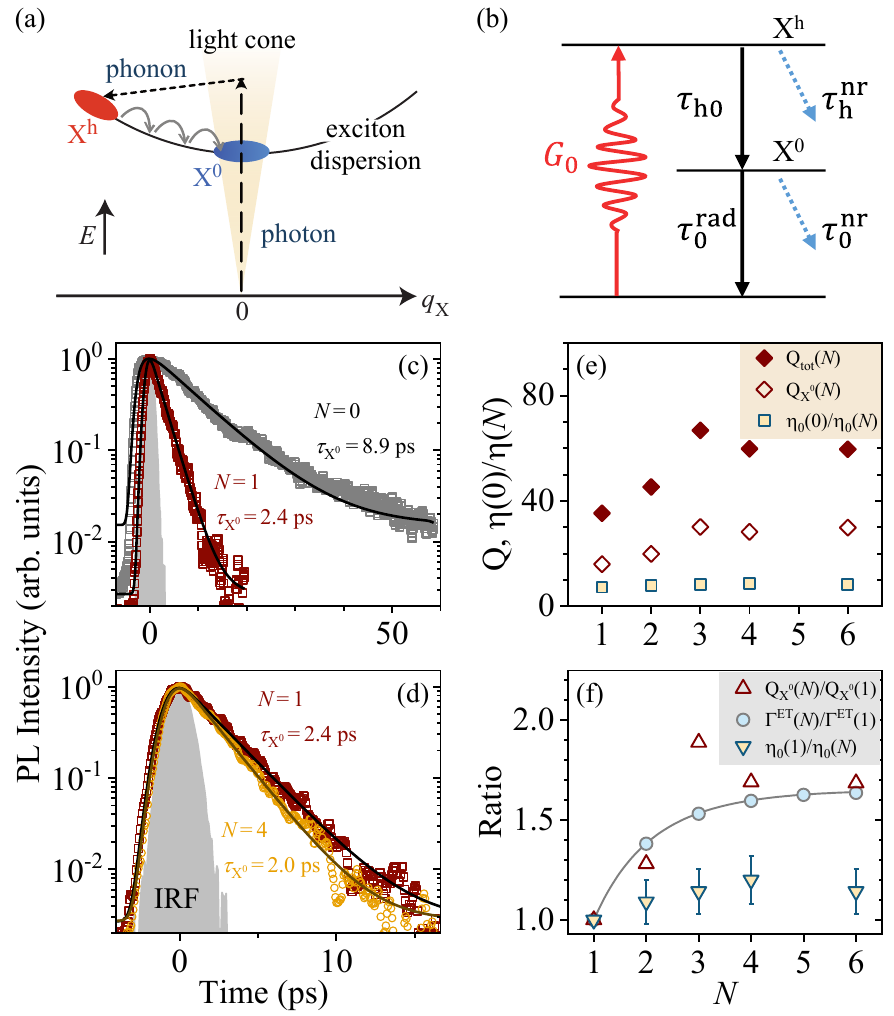}
    \caption{(a) Sketch of exciton formation in the momentum $\left (q_{\mathrm X}\right ) -$ energy $(E)$ plane. (b) Corresponding three-level system and characteristic times. Hot excitons ($\mathrm X^{\mathrm h}$)  with finite center of mass momentum $q_{\mathrm X}$ are pumped non-resonantly (through phonon-assisted processes) by ultrashort laser pulses and may relax into cold excitons with near-zero momentum ($\mathrm X^0$), which can radiatively recombine, with a decay time $\tau_0^{\mathrm{rad}}$. Non-radiative losses are considered for hot and cold excitons, with decay times $\tau_{\mathrm {h}}^{\mathrm{nr}}$ and $\tau_{\mathrm {0}}^{\mathrm{nr}}$, respectively. Graphene opens non-radiative transfer channels that shorten $\tau_{\mathrm {h}}^{\mathrm{nr}}$ and $\tau_{\mathrm {0}}^{\mathrm{nr}}$. (c) Picosecond TRPL of the MoSe$_2$ reference ($N=0$) and of hBN/MoSe$_2/1$LG ($N=1$) measured at 6~K on Sample 1. The solid lines are fits to the data using the three-level system in (b) and considering a convolution with the instrument response function (IRF, gray area).  The PL rise is unresolved and the $\mathrm X^0$ decay times $\tau_{X^0}$ are indicated. (d) Comparison between the TRPL of MoSe$_2/1$LG ($N=1$) and MoSe$_2/4$LG ($N=4$) measured on Sample 1. (e) The PL quenching factors $Q_{\mathrm{tot}}(N)$ and $Q_{\mathrm{X^0}}(N)$ extracted from PL spectra measured in the same conditions and on the same spots as the TRPL data are compared to $\eta(0)/\eta(N)$, the ratio between the emission yield of the MoSe$_2$ reference and that of MoSe$_2/N$LG, deduced from the TRPL measurements. (f) An analytical calculation of the normalized F\"orster-type energy transfer (FRET) rate $\Gamma^{\mathrm{ET}}\left(N\right)/\Gamma^{\mathrm{ET}}\left(1\right)$ is compared to our measurements of the $\mathrm X^0$ emission yield ratio $\eta_0(1)/\eta_0(N)$ and to the increase of $Q_{\mathrm{X^0}}(N)$ relative to $Q_{\mathrm{X^0}}(1)$. TRPL measurements were performed in the linear regime, with femtosecond laser pulses at 1.71~eV and a pulse fluence below $10\:\mu\mathrm J/\mathrm{cm}^2$. The values of $\tau_{\mathrm X^0}$ are obtained with less than $10\%$ uncertainty. Error bars are shown only when they exceed the symbol size in (e,f).} 
    \label{Fig_S3_TRPL}
\end{figure}

The PL spectrum of bare MoSe$_2$ (Fig.~\ref{Fig1}a,d) is dominated by two narrow lines of a few meV in full-width at half maximum (FWHM)~\cite{SMnote} arising from bright neutral excitons ($\mathrm X^0$ near $1.65~\mathrm{eV}$) and charged excitons (trions, $\mathrm X^{\star}$, near  $1.62~\mathrm{eV}$). The most striking effect of a tightly coupled graphene layer is to filter the MoSe$_2$ PL spectrum (Fig.~\ref{Fig1}b,c), leaving one single emission line arising from $\mathrm X^0$~\cite{Lorchat2020,Parralopez2021}. The intensity of the $\mathrm X^0$ PL line is quenched by more than one order of magnitude in MoSe$_2/N$LG with respect to the MoSe$_2$ reference. Interestingly, the quenching factor only augments weakly but still appreciably with increasing $N$ (Fig.~\ref{Fig1}b,c, Fig.~\ref{Fig_S3_TRPL}e and Supplemental Material~\cite{SMnote}). Due to dielectric screening~\cite{Raja2017,Lorchat2020}, the $\mathrm X^0$ line is slightly redshifted by about 10~meV in MoSe$_2$/$N$LG with fluctuations due to local variations of built-in strain, electrostatic environment and dielectric screening~\cite{Raja2019}. The spatially averaged shift does not significantly depend on the number of graphene layers suggesting that dielectric screening is mostly determined by the first coupled atomic layer~\cite{SMnote}. 
Remarkably, an ultrathin hBN spacer (2 and 5 layers, with thicknesses of about 0.7~nm and 1.7~nm, respectively, see Fig.~\ref{Fig1}e and f, respectively)  between MoSe$_2$ and graphene suffices to decouple MoSe$_2$ from graphene, such that no significant PL quenching is observed with respect to the MoSe$_2$ reference. We note that the $\mathrm X^{\star}$ spectral weight is reduced in MoSe$_2/N$L\:hBN/1LG as compared to the MoSe$_2$ reference, likely because the proximal graphene layer provides a more homogeneous electrostatic environment.

~


\textbf{TRPL on MoSe$_2/N$LG -- }Figure~\ref{Fig_S3_TRPL} shows typical TRPL measurements performed on Sample 1 (see Fig.~\ref{Fig1}a-c)) using a streak camera~\cite{SMnote}. The high resolution ($\approx 1\mathrm ps$) of these measurements enable us to determine $\tau_{\mathrm{X^0}}$, the $\mathrm X^0$ lifetime in MoSe$_2/N$LG.

Let us first consider the SiO$_2$/MoSe$_2$/hBN reference (see Fig.~\ref{Fig_S3_TRPL}c). The measured TRPL reveals a sub-picosecond rise time, i.e., faster than the instrument response function and a decay time of $(8.9\pm 0.5)~\mathrm {ps}$. The decay time is assigned to the relaxation of cold excitons ($\mathrm X^0$ in Fig.~\ref{Fig_S3_TRPL}a,b), that is maximized here due cavity effects~\cite{Fang2019,SMnote}. The sub-picosecond rise time is similar to previous measurements on SiO$_2$ supported MoSe$_2$~\cite{Robert2016} and assigned to the lifetime of the hot exciton reservoir ($\mathrm X^\mathrm h$, see Fig.~\ref{Fig_S3_TRPL}a,b) formed almost instantaneously after non-resonant photoexcitation well below the free-carrier continuum~\cite{Mourzidis2025} and from which $\mathrm X^0$ are formed.

Non-radiative $\mathrm X^0$ relaxation to graphene leads to shorter PL decays in MoSe$_2/N$LG ($\gtrsim 2\:\mathrm{ps}$ (Fig.~\ref{Fig_S3_TRPL}c,d and Supplemental Material~\cite{SMnote}), while the PL rise time remains below our resolution limit. Interestingly,  the decay time measured in MoSe$_2/N$LG  is nearly independent of $N$ and we only observe a marginally longer decay time by less than 20$\%$ in MoSe$_2/1$LG as compared to MoSe$_2/N$LG $(N>1)$, a value that is slightly above our experimental error, conservatively estimated below $10~\%$. These results confirm that $\mathrm X^0$ relaxation in MoSe$_2/N$LG is essentially determined by the first graphene layer.

Considering the reduction of the exciton binding energy in MoSe$_2/N$LG, which we estimate through hot PL measurements of the excited excitonic states~\cite{Lorchat2020}, we determine that the $\mathrm X^0$ radiative lifetime in MoSe$_2/N$LG is approximately 2 times larger than in the MoSe$_2$ reference and that this increase barely depends on $N$. Consequently, the measured $\tau_{\mathrm{X^0}}$ are chiefly determined by the transfer times to graphene that we estimate from $2.8~\mathrm {ps}$ in MoSe$_2/1$LG, down to $2.2-2.4 ~\mathrm {ps}$ in MoSe$_2/N$LG. Extensive PL and TRPL measurements were also performed on a similar sample and we found consistent results, notably an exciton transfer time within $10\%$ of the values discussed above (see Sample S4~\cite{SMnote}).

~

\begin{figure}[t!]
    \centering
    \includegraphics[width=1\linewidth]{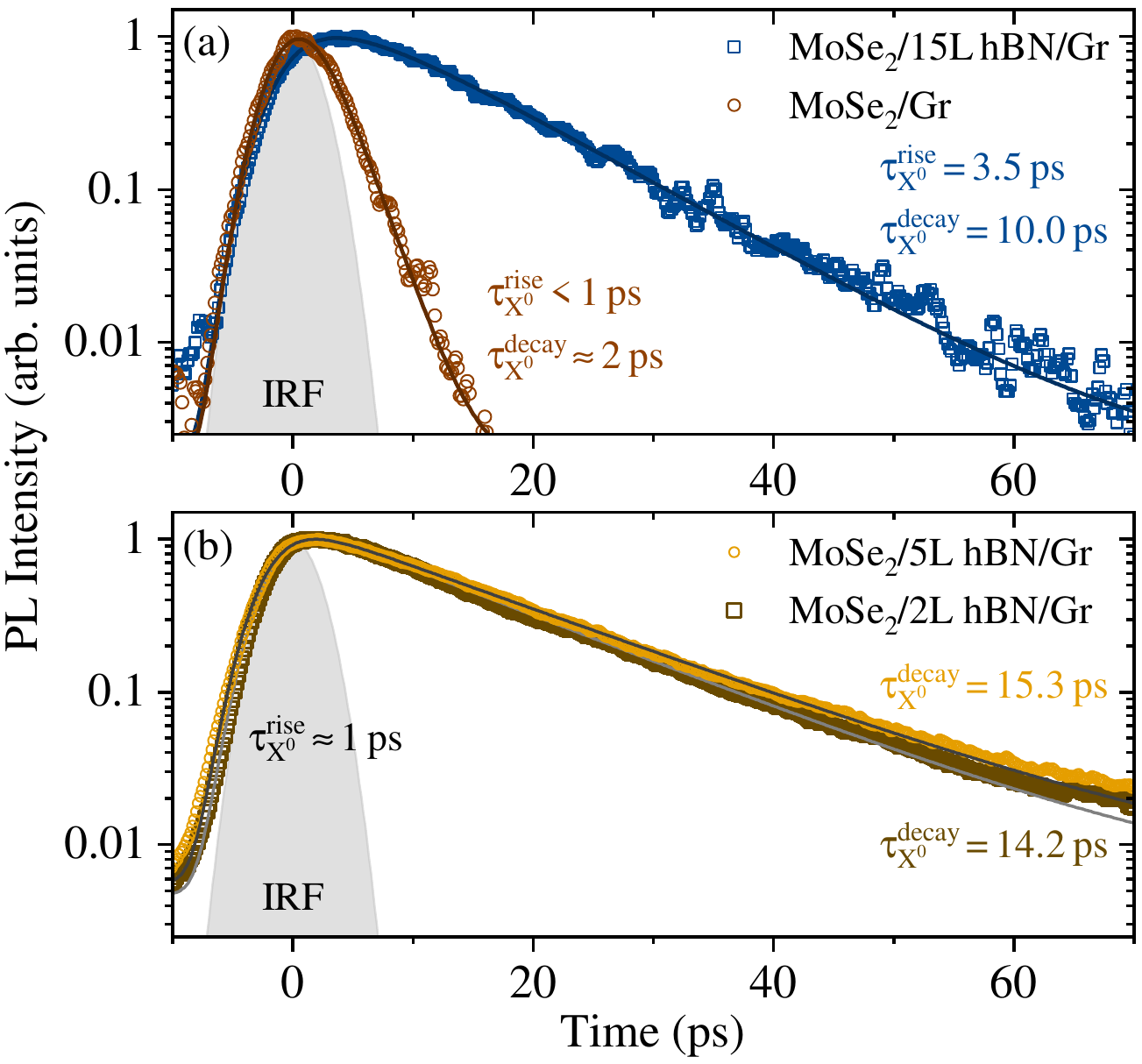}
    \caption{TRPL measurements on the $\mathrm X^0$ line in (a) Sample 3 with hBN-encapsulated MoSe$_2$/15L\:hBN/1LG and MoSe$_2$/1LG and (b) Sample 2 with hBN-encapsulated MoSe$_2$/2L\:hBN/1LG and MoSe$_2$/5L\:hBN/1LG. Solid lines are fit to the data as in Fig.~\ref{Fig_S3_TRPL}. Measurements were performed in the linear excitation regime at 6~K under pulsed laser excitation (pulse duration $<2\:$ps) at 1.73~eV.} 
    \label{Fig_iBN_TRPL}
\end{figure}

\textbf{TRPL on MoSe$_2/N\mathbf{L\:hBN/1LG}$ -- } We now discuss the TRPL of Sample 3, which displays MoSe$_2$/1LG MoSe$_2/15\mathrm {L\:hBN/1LG}$ domains (Fig.~\ref{Fig_iBN_TRPL}a), as well as of Sample 2, introduced in Fig.~\ref{Fig1}d-f (see Fig.~\ref{Fig_iBN_TRPL}b).  Considering the thickness of the bottom hBN flakes used to fabricate Samples 2 and 3, we estimate that the MoSe$_2$ layer is quite close to a node of the optical field in Sample 3 (as in Sample 1), while it is approaching an anti-node in Sample 2~\cite{Fang2019,SMnote}. 
We first consider Sample 3, where the MoSe$_2/1$LG domain displays a short $\mathrm X^0$ decay within about 2~ps, very similar to Sample 1 (Fig~\ref{Fig_S3_TRPL}c,d). On the same sample, when the graphene layer is decoupled by a 15~L hBN spacer, we observe much slower dynamics with rise and decay times of $3.5\pm 0.5 ~\mathrm {ps}$ and $10.0\pm 0.5~\mathrm{ps}$. On Sample 2, with 2L\:hBN- and 5L\:hBN-thick spacers, we observe very similar TRPL traces, irrespective of the spacer thickness, with a short $\approx 1\:\mathrm {ps}$ rise time that we cannot resolve and a longer decay in $15\pm 1~\mathrm{ps}$. The TRPL traces in the MoSe$_2/N\mathrm{L\:hBN/1LG}$ domains of Samples 2 and 3 are very similar to previous reports in hBN-encapsulated MoSe$_2$ within similar optical and dielectric environments~\cite{Fang2019,Venanzi2024,Lorchat2020}. In keeping with Ref.~\cite{Fang2019}, we assign the PL rise time in Samples 2 and 3 to $\tau_{\mathrm{X^0}}$ and the PL decay time to the $\mathrm {X^h}$ relaxation time, respectively, and we conclude that the low temperature $\mathrm X^0$ dynamics of MoSe$_2$ are not appreciably affected by graphene as soon as the MoSe$_2$-graphene distance exceeds one atomic layer of hBN.

~

\textbf{Discussion: cold excitons ($\mathrm X^0$) -- }
We now comment on the microscopic mechanisms leading to a reduction of the $\tau_{\mathrm{X^0}}$ and $\mathrm X^0$ PL quenching. As discussed previously, both short-range charge tunelling and FRET~\cite{Forster1948}  may contribute. FRET from TMD excitons to graphene has recently been considered theoretically~\cite{Selig2019} and experimentally~\cite{Froehlicher2018,Lorchat2020,Dong2023,Tebbe2024}, following numerous studies in 2D-2D hybrid systems, including organic/inorganic semiconductor heterostructures~\cite{Basko1999} and semiconductor nanoplatelet/graphene heterostructures~\cite{Federspiel2015}. In contrast to the case of point-like systems or $\mathrm{0D-2D}$ heterostructures~\cite{Chen2010,Gaudreau2013,Federspiel2015,Prins2014}, conservation of the exciton center of mass momentum in a 2D-2D system has major consequences on the FRET rate. Indeed, since the momentum-dependent FRET efficiency scales as $q_{\mathrm X}^2\exp{\left(-2q_{\mathrm X}d\right)}$~\cite{Basko1999,Selig2019}, where $d$ is the TMD-graphene distance, no FRET is expected for excitons at $q_{\mathrm X}=0$ and more broadly FRET is inefficient for $\mathrm X^0$ excitons in the light cone (such that $q_{\mathrm X}\lesssim 10^{7} \mathrm m^{-1}$), i.e., $q_{\mathrm X}\ll E_{\mathrm X^0}/\left (\hbar v_{\mathrm F}\right )$, with $E_{\mathrm X^0}$ the $\mathrm{X}^0$ energy and $v_{\mathrm F}$ the  Fermi velocity of graphene.

In addition, at cryogenic temperature the FRET rate is nearly constant for $d\lesssim 1~nm$ and then follows a smooth exponential decay with a characteristic length in the few nanometer range. For $d\gtrsim 5~\mathrm{nm}$, the well-known $1/d^4$ scaling is expected.  These theoretical predictions~\cite{Selig2019,SMnote} contrast with the drastically different (TR)PL characteristics measured in MoSe$_2/1$LG (Fig.~\ref{Fig1}b,~\ref{Fig_S3_TRPL}c,~\ref{Fig_iBN_TRPL}a) and MoSe$_2/2\mathrm{L\:hBN/1LG}$ (Fig.~\ref{Fig1}e,~\ref{Fig_iBN_TRPL}b), as well as with our conclusion that a 2-layer thick hBN spacer (i.e., 0.7~nm) is sufficient to recover (TR)PL features akin to bare MoSe$_2$. Studies at room temperature, where much more efficient FRET is expected for $\mathrm{X^0}$, have revealed that the distance-dependence of the transfer rate deviates strongly from theoretical predictions based solely on FRET, when the TMD-graphene distance is within the tunneling range (that is, with no spacer between TMD and graphene or just one atomic layer of hBN)~\cite{Tebbe2024}.

We thus conclude that another microscopic mechanism must be at play in order to explain the reduction of $\tau_{\mathrm{X^0}}$ that we consistently observe in MoSe$_2/N$LG. Considering that the $\mathrm X^0$ dynamics in MoSe$_2$/graphene systems i) are drastically modified by the presence of sub-nm thick dielectric spacer of hBN (Fig.~\ref{Fig_iBN_TRPL}) and then marginally depend on the spacer thickess, and ii) depend very weakly on the number of graphene layers, it is sensible to invoke charge tunneling. Dexter energy transfer (also known as electron exchange)~\cite{Dexter1953} may in principle contribute but the predicted rate in TMD/graphene is orders of magnitude smaller than the FRET rate, chiefly due to limited overlap between TMD and graphene electronic orbitals~\cite{Dong2023}. The only plausible mechanism that may account for the shortening of $\tau_{\mathrm{X^0}}$ and its dependence on $d$ and on the number of graphene layers is direct charge tunneling of the electrons and holes that compose $\mathrm X^0$. Since our previous experiments did not reveal any fingerprint of a long-lived charge separated state~\cite{Aeschlimann2020} on timescales on the order of $\tau_{\mathrm{X^0}}$~\cite{Ferrante2022}, we propose that sequential (balanced) tunneling of electrons and holes is responsible for the shortening of $\tau_{\mathrm{X^0}}$.

\textbf{Discussion: hot excitons ($\mathrm {X^h}$) --} These finite momentum excitons can be transferred to graphene before forming $\mathrm X^0$, thus contributing to the measured PL quenching~\cite{SMnote}. Along these lines, we compare, in Fig.~\ref{Fig_S3_TRPL}e, $\eta(0)/\eta(N)$, the ratio between the emission yield of the MoSe$_2$ reference and that of MoSe$_2/N$LG domains, deduced from TRPL measurements to the $\mathrm X^0$ PL quenching factor $Q_{\mathrm X^0}(N)=I_{\mathrm X^0}(0)/I_{\mathrm X^0}(N)$ and to the total PL quenching factor $Q_{\mathrm {tot}}(N)=I_{\mathrm {tot}}(0)/I_{\mathrm {tot}}(N)$, all recorded simultaneously during TRPL measurements on Sample 1. Here, $I_{\mathrm X^0}$ and $I_{\mathrm {tot}}$ denote the integrated PL intensity of the $\mathrm X^0$ line and the total PL intensity (considering $\mathrm X^0$ and $\mathrm X^{\star}$ emission), respectively, while  $(0)$ and $(N)$ refer to the MoSe$_2$ reference and the MoSe$_2/N$LG domains, respectively.

With the three level system in Fig.~\ref{Fig_S3_TRPL}b, assuming identical $\mathrm{X^h}$ generation rates in MoSe$_2$ and MoSe$_2/N$LG, we obtain $Q_{\mathrm X^0}(N)=\frac{\eta_{\mathrm F}(0)}{\eta_{\mathrm F}(N)}\frac{\eta_{0}(0)}{\eta_{0}(N)}$, where $\eta_{\mathrm F}$ is the $\mathrm X^0$ formation yield~\cite{SMnote}. Since $\mathrm X^{\star}$ emission can be neglected in MoSe$_2/$Gr, we get $Q_{\mathrm {tot}}(N)=Q_{\mathrm X^0}(N)\left(1+\frac{I_{\mathrm X^{\star}}(0)}{I_{\mathrm X^0}(0)}\right)$.
Based on the lifetimes measured in Fig.~\ref{Fig_S3_TRPL} and on our estimation of the two-fold increase of the radiative lifetime in MoSe$_2/N$LG~\cite{Lorchat2020}, we get $\frac{\eta_{0}(0)}{\eta_{0}(N)}\approx 8$ for Sample 1, a value that is nearly independent on $N$ and largely smaller than the values of $Q_{\mathrm {X_0}}$ (resp. $Q_{\mathrm{tot}}$), which range from  16 (resp. 35) for $N=1$ to 30 (resp. 60) for $N>2$ (see Fig.~\ref{Fig_S3_TRPL}e). 

We can thus estimate that the $\mathrm X^0$ formation yield is reduced by a factor ranging from approximately 2 to 4 as $N$ increases. More broadly, we have systematically observed on all the samples studied in this work that the measured quenching factors $Q_{\mathrm tot}$ and $Q_{\mathrm X^0}$ are larger than $\eta_0(0)/\eta_0(N)$~\cite{SMnote}. 
This discrepancy is assigned to fast (sub-ps and unresolved here) transfer of $\mathrm{X^h}$.  In this regard, it is interesting to compare  the evolution of the normalized FRET rate ($\Gamma^{\mathrm{ET}}(N)/\Gamma^{\mathrm{ET}}(1)$) calculated considering contributions from all possible $q_{\mathrm X}$ to the quenching factor ratio $Q_{\mathrm{X^0}}(N)/Q_{\mathrm{X^0}}(1)$ and the emission yield ratio $\eta_0(1)/\eta_0(N)$~\cite{SMnote}. As expected, the latter ratio remains close to unity whereas we observe a good qualitative agreement between the predicted evolution of the FRET rate and of the quenching factor with $N$, as shown in Fig.~\ref{Fig_S3_TRPL}f. This result confirms that FRET alone cannot account for the accelerated relaxation of $\mathrm X^0$ but that FRET may contribute substantially to the sub-ps transfer of $\mathrm {X^h}$ to graphene and subsequent $\mathrm X^0$ PL quenching.

We finally note that an alternate mechanism, involving intraband transitions in graphene enabled by a transient population of hot holes near the Fermi level of graphene, and referred to as Meitner Auger energy transfer has been proposed to account for sub-ps exciton transfer in WSe$_2$/graphene under intense pulsed laser excitation~\cite{Dong2023}. Importantly, this mechanism involves excitons  with large $q_{\mathrm X}\approx E_{\mathrm {ph}}/(\hbar v_{\mathrm F})\approx 2.5 \times 10^9 \mathrm\: m^{-1}$,  with  $E_{\mathrm {ph}}$ the photon energy and does not affect $\mathrm X^0$. Meitner-Auger transfer might affect $\mathrm {X^h}$, however, our measurements are performed at pulse fluences that are more than two orders of magnitude lower than in Ref.~\cite{Dong2023}, such that this transfer mechanism can safely be neglected in first approximation.

\textbf{Conclusion and outlook --}
The low temperature (typically $T\leq16~\mathrm K$) relaxation of the lower lying, bright excitons in MoSe$_2/N$-layer graphene is largely independent on $N$ and determined by sub-nm range picosecond charge tunneling to graphene. Slower exciton dynamics, typical from a bare MoSe$_2$ monolayer, are recovered for separations in excess of one atomic layer between MoSe$_2$ and graphene. These results demonstrate that F\"orster-type resonant energy transfer is inefficient for cold, bright excitons at low temperatures,  as the center of mass momentum of these excitons is vanishingly small, but may contribute significantly to the transfer of hot, finite momentum excitons. This hypothesis is supported by the increase of the PL quenching factor with $N$, in agreement with an electrodynamical model.

Exciton dynamics are known to be complex in bare TMD monolayers due to a subtle interplay between bright exciton formation and recombination kinetics, as well as a sizeable contribution from trions and, more broadly, spatial heterogeneities in the dielectric and electrostatic landscape. This situation leads to challenging assignments of photoluminescence rise and decay times~\cite{Robert2016,Lorchat2020,Venanzi2024}. In contrast, we observe particularly robust and reproducible bright exciton photoluminescence dynamics in TMD/graphene, with a sub-ps rise time and a decay time in the range $2.0 - 2.5\:\mathrm {ps}$, that are largely independent of the subtleties that affect exciton dynamics in the TMD reference.

Our work invites further experimental and theoretical investigations of charge tunneling and energy transfer involving other excitonic species such as spin- or momentum-dark excitons, interlayer excitons, including moiré-trapped or -hybridized excitons, and more broadly in proximitized van der Waals materials~\cite{Wang2018,Wilson2021,Zutic2019}.

\begin{acknowledgements}

We are grateful to the STnano cleanroom staff for technical support. We acknowledge financial support from the Agence Nationale de la Recherche (under grants ATOEMS ANR-20-CE24-0010, VANDAMME ANR-21-CE09-0022, TEXTURES ANR-22-CE09-0008, INFERNO ANR-22-CE42-0015, EXODUS ANR-23-QUAC-0004, as well as under the program ESR/EquipEx+ (ANR-21-ESRE- 0025). This work of the Interdisciplinary Thematic Institute QMat, as part of the ITI 2021 2028 program of the University of Strasbourg, CNRS and Inserm, was supported by IdEx Unistra (ANR 10 IDEX 0002), and by SFRI STRAT'US project (ANR 20 SFRI 0012) and EUR QMAT ANR-17-EURE-0024 under the framework of the French Investments for the Future Program.  S.B. and A.R.M.  acknowledge support from the Indo-French Centre for the Promotion of Advanced Research (CEFIPRA). S.P. acknowledges support from the Swiss National Science Foundation (SNSF-grant no. 230675). K. W. and T. T. acknowledge support from the Elemental Strategy Initiative conducted by the MEXT, Japan, and the CREST (JPMJCR15F3), JST.

\end{acknowledgements}


%




\onecolumngrid
\newpage
\begin{center}
{\Large\textbf{Supplemental Material for:\\ Sub-nm range momentum-dependent exciton transfer\\ from a 2D semiconductor to graphene}}
\end{center}

\setcounter{equation}{0}
\setcounter{figure}{0}
\setcounter{section}{0}
\renewcommand{\thetable}{S\arabic{table}}
\renewcommand{\theequation}{S\arabic{equation}}
\renewcommand{\thefigure}{S\arabic{figure}}
\renewcommand{\thesection}{S\arabic{section}}
\renewcommand{\thesubsection}{S\arabic{section}.\arabic{subsection}}
\renewcommand{\thesubsubsection}{\textbf{S\arabic{section}.\arabic{subsection}.\arabic{subsubsection}}}

\linespread{1.4}\selectfont

\bigskip



\section{\textbf{Experimental methods}}
\label{Methods}

\begin{figure}[ht!]
    \begin{center}
    \includegraphics[width=1.0\linewidth]{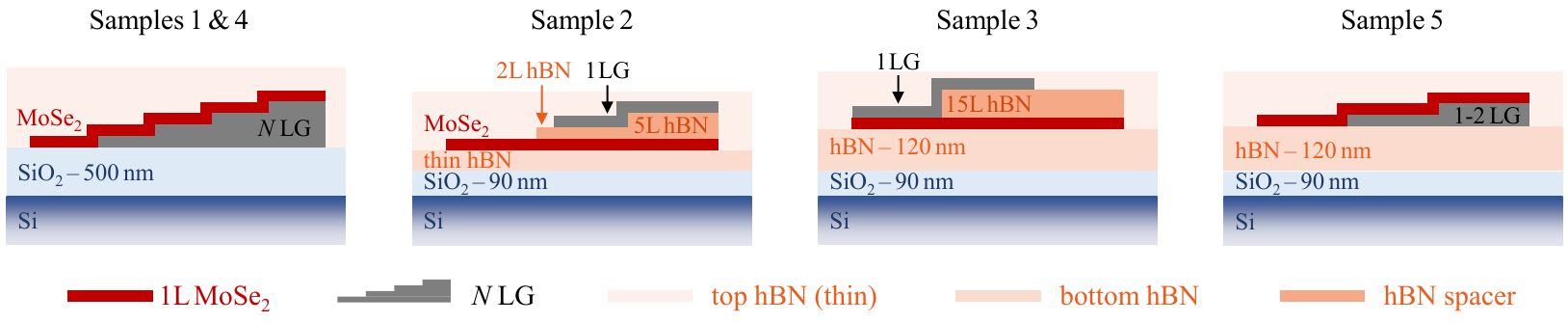}
    \caption{\textbf{Samples --} Inventory of the van der Waals heterostructures mentioned in our work.}
    \label{Fig_SM_samples}
    \end{center}
\end{figure}

\textbf{Fabrication of van der Waals heterostructures --} 
The van der Waals heterostructures discussed in this study are skteched in Fig.~\ref{Fig_SM_samples} and are numbered by order of appearance in our main manuscript. Our samples fall into two categories. First, ``straircase-like'' MoSe$_2/N$-layer graphene samples, such as Samples 1, 4 and 5; second MoSe$_2/N$-layer hBN/1-layer graphene samples, where the graphene monolayer is decoupled from the MoSe$_2$ monolayer by a thin spacer of hexagonal boron nitride (hBN), such as samples 2 and 3.
All samples  were made by mechanical exfoliation of bulk crystals (graphite, hBN, MoSe$_2$), followed by the ``pick-up and lift'' method, using an additional polycarbonate (PC) stamp that is later dissolved in chloroform~\cite{Wang2013,Zomer2014}. More details on the sample geometry and associated optical cavity effects are discussed in Sec.~\ref{SecCav}.


\textbf{Photoluminescence (PL) spectroscopy and PL mapping --} The steady state PL spectroscopy measurements were performed at cryogenic temperature in a continuous flow optical cryostat. Continuous wave (cw) laser excitation at 1.96~eV and 2.33~eV with low intensity below 25 $\mu \mathrm {W/\mu m}^2$ was employed, in order to remain in the linear excitation regime. The photoluminescence excitation spectroscopy (PLE) measurements were done by measuring the PL spectra  at $\sim 16~\mathrm K$ with tuning the incident laser photon energy from 1.69~eV to 2.57~eV using a supercontinuum laser (see Fig.~\ref{Fig_S3_PLE_spectra},~\ref{Fig_S3_PLE} and ~\ref{Fig_S3_PLE_FWHM} in Sec.~\ref{SectionData}.1.5 for Sample 1 and  Fig.~\ref{Fig_S8_PLE} in Sec.~\ref{SectionData}.4 for Sample 4). Typical laser intensities around $10\:\mu \mathrm {W/\mu m}^2$ were used at each wavelength. 

~

\textbf{Time-resolved PL spectroscopy (TRPL) --} Time-resolved PL measurements were performed at $6~\mathrm K$ in a closed-cycle optical cryostat using a synchro-scan streak camera with a temporal resolution near 1.0~ps for Sample 1 (Fig. 2) and Sample 5 (Fig.~\ref{Fig_Luis_PL_TRPL}) and near 2.5~ps for Samples 2 and 3 (Fig. 3). An oscillator delivering $\approx 150~\mathrm fs$ pulses at an energy $1.71~\mathrm{eV}$ (measurements on Samples 1 and 5) and $< 2~\mathrm ps$ pulses at 1.73~eV (measurements on Samples 2 and 3). Noteworthy, in both bases, the laser photon energy was below the first excited excitonic state ($\mathrm X^0_{2\mathrm s}$)~\cite{Wang2018} and the pulse duration was shorter than the temporal resolution of the streak camera. The average laser intensity at the sample was maintained low, typically a few $\mu \mathrm {W/\mu m}^2$, to ensure that our studies were performed in the linear excitation regime. The TRPL data were fit using the convolution of the instrument response function (IRF, fit to an hyperbolic secant) and Eq.~\eqref{EqTRPL} (see Sec.~\ref{SecRate}).

~

\section{\textbf{Sample design and exciton lifetime engineering}}
\label{SecCav}

\begin{figure}[ht!]
    \begin{center}
    \includegraphics[width=0.6\linewidth]{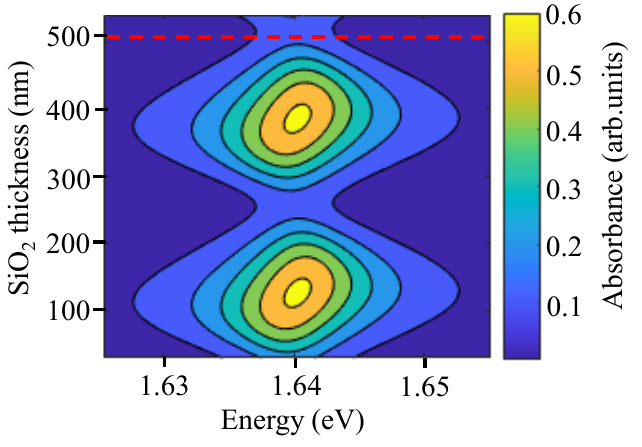}
    \caption{\textbf{Cavity engineering --} Calculated absorbance (in arbitrary units) by an MoSe$_2$ monolayer sandwiched between the SiO$_2$ epilayer of an Si/SiO$_2$ substrate and thin (typically 10-nm thick) hBN top layer as a function of SiO2$_2$ thickness and laser photon energy. The $\rm X^0$ excitonic resonance near 1.64~eV is clearly visible. The absorbance has a minimum for a 500~nm SiO$_2$ thickness (as is the case for Samples 1 and 4), which translates into a maximum in the radiative lifetime as confirmed by our TRPL measurements in Fig.~2c and in Fig.~\ref{Fig_Luis_PL_TRPL}b.}
    \label{Fig_cavity}
    \end{center}
\end{figure}

All our van der Waals heterostructures were placed in a layered, cavity-like, environment, in which we could engineer the radiative lifetime of MoSe$_2$ excitons ($\mathrm X^0$) as discussed extensively in \cite{Fang2019}. In brief, this radiative lifetime depends directly on the local density of optical states at the position of the MoSe$_2$ monolayer within the cavity. As an example, in Figure~\ref{Fig_cavity}, we show a calculation of the absorbed power by an MoSe$_2$ monolayer, within the geometry of Samples 1 and 4. Simulations were made using a transfer matrix approach as in Ref.~\cite{Fang2019}. It is clear that in this geometry, the MoSe$_2$ layer is placed near a node of the electromagnetic field, such that the $\mathrm X^0$ radiative lifetime is increased as compared to the typical values of 2~ps observed on transparent substrates without cavity effects~\cite{Lorchat2020}. 

The other samples use a 90~nm-thick SiO$_2$ layer and a thin (typically 10~nm to 15~nm) hBN top layer. In this case, it was previously reported that the radiative lifetime reaches a minimum for a thin bottom hBN layer (below 20~nm) and a maximum for an bottom hBN thickness near 120~nm~\cite{Fang2019}. We chose a thin bottom hBN layer for Sample 2 and an $\approx\:120\:\mathrm{nm}$-thick bottom hBN layer for Sample 3 and Sample 5.

\begin{figure}[ht!]
    \begin{center}
    \includegraphics[width=0.8\linewidth]{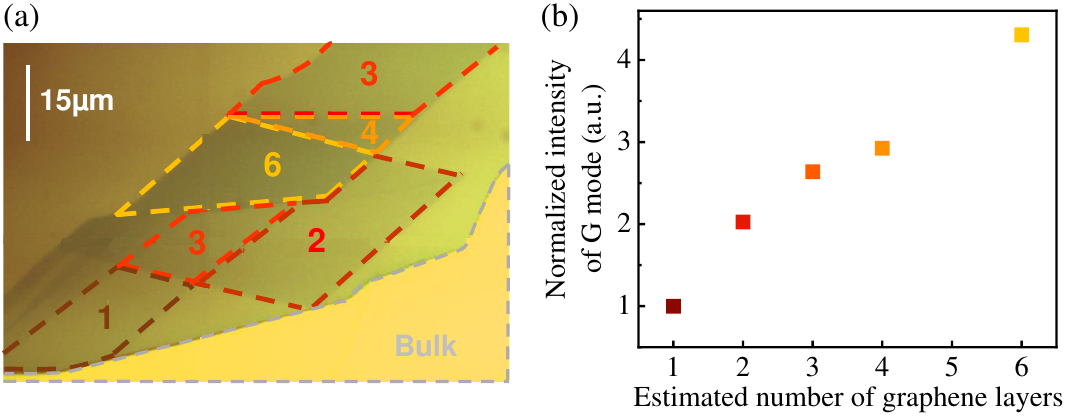}
    \caption{\textbf{Estimation of the number of layers in Sample 1 --} (a) Optical micrograph of the $N-$layer graphene flake used to make Sample 1 (see Fig.~1 and Fig.~2) in the main manuscript. (b) Integrated intensity of the Raman G-mode in the different regions of the sample. Due to optical interference effects the scaling with $N$ is expected to deviate slightly from linearity but still allows us to estimate $N$ with confidence.}
    \label{fig_SM_GrThickness}
    \end{center}
\end{figure}

The thickness of the graphene layers was determined by Raman spectroscopy (see Sec.~\ref{SectionData}.1 and Fig.~\ref{fig_SM_GrThickness}). The thickness of the ``top'' and ``bottom'' hBN layers used for samples was estimated by examining the optical contrast before pick-up~\cite{Fang2019}. The thickness of the ultrathin hBN spacers used in Samples 2 and 3 (see Fig. 1d-f and Fig.~3 in the main text) was determined by atomic force microscopy measurements as discussed in Sec.~\ref{SectionData}.2 and ~\ref{SectionData}.3  below.

\clearpage

\section{\textbf{Supplementary data}}
\label{SectionData}

\subsection{\textbf{Sample 1}}

\subsubsection{\textbf{Determining the number of graphene layers}}

\clearpage

\subsubsection{\textbf{PL spectroscopy and hyperspectral mapping on Sample 1}}

\begin{figure}[ht!]
    \centering
    \includegraphics[width=1\linewidth]{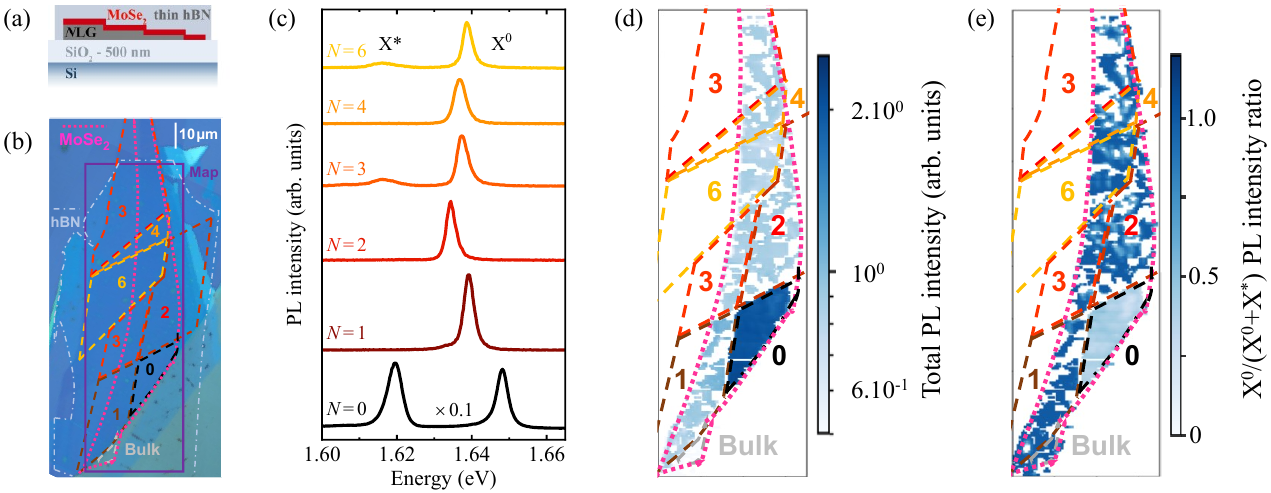}
    \caption{\textbf{Photoluminescence mapping -- }
(a) Schematic of Sample 1. (b) Optical image of Sample 1 with the dashed, dotted and dash-dotted contours depicting areas with $N$-layer graphene, MoSe$_2$ and hBN respectively. The solid rectangle shows the area on where the PL map was recorded. (c) PL spectra taken at temperature of 16~K with laser excitation from 1.96 eV laser and intensity close to 25 $\mu$W/$\mu$m$^2$, denoting the bright exciton $(\rm X^0)$ and trion $(\rm X^{\star})$ peaks. The extracted parameters from the PL map over the mapped area with dashed contours denoting $N$-layer graphene: (d) total PL intensity and (e) ratio of intensity of $\rm X^0$ to total PL intensity.} 
    \label{FigSM_S3Id}
\end{figure}

\begin{figure}[htb!]
    \begin{center}
    \includegraphics[width=0.7\linewidth]{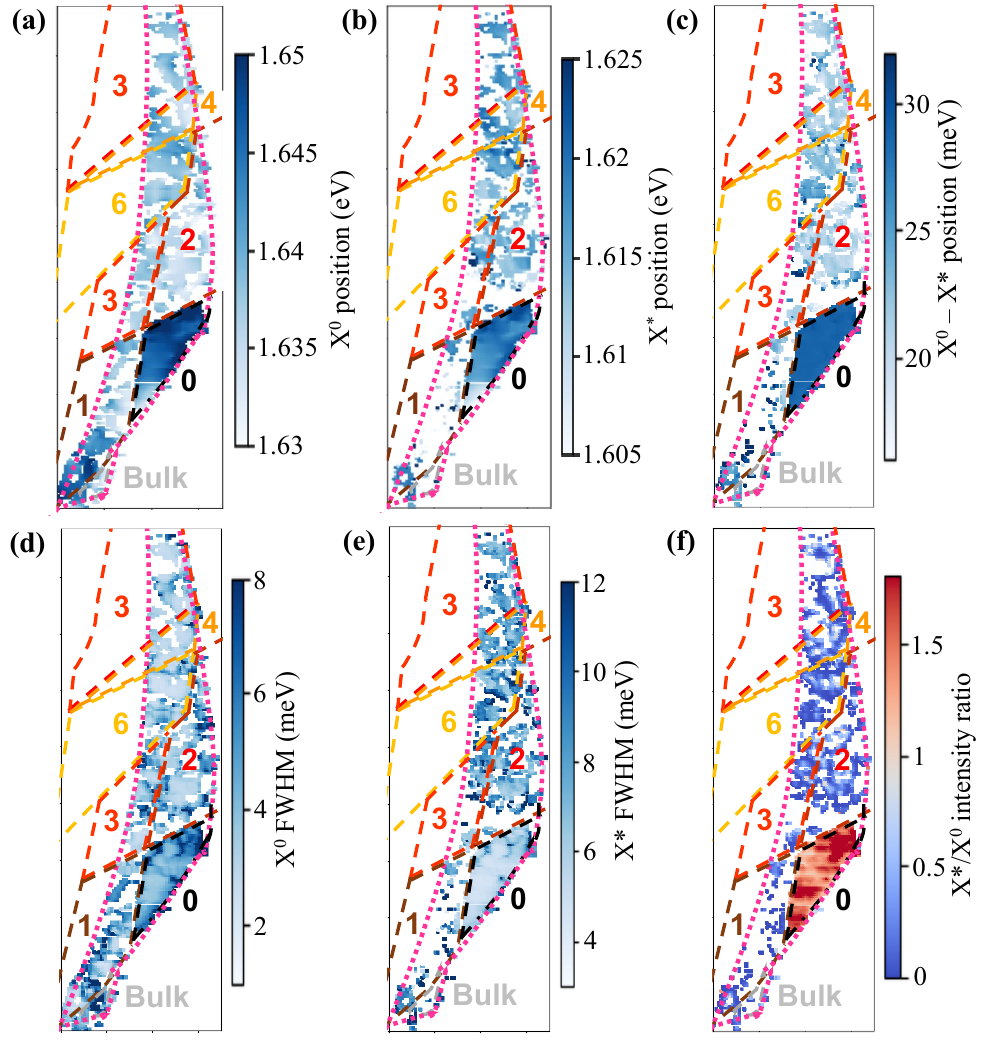}
    \caption{\textbf{PL mapping of Sample 1 --} Hyperspectral maps of (a) bright exciton ($\rm X^0$) energy (b) Trion ($\rm X^{\star}$) energy (c) Energy difference between bright exciton and trion energy, i.e., trion binding energy, (d) FWHM of the $\rm X^0$ photoluminescence (PL) feature (e) FWHM of the $\rm X^{\star}$ feature (f) Trion-to-exciton integrated intensity ratio.}
    \label{fig_SI_S3_Map}
    \end{center}
\end{figure}

\begin{figure}[ht!]
    \centering
    \includegraphics[width=0.6\linewidth]{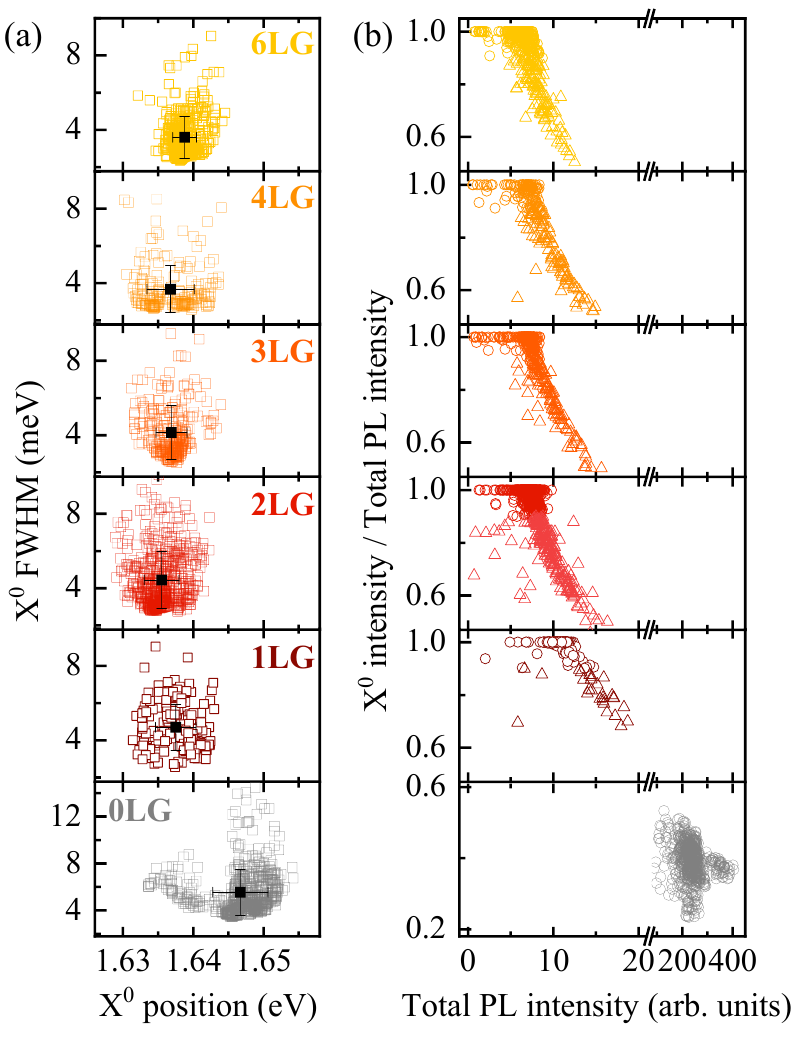}
    \caption{\textbf{Correlation plots from PL mapping of Sample 1 (see Fig.~\ref{FigSM_S3Id} and \ref{fig_SI_S3_Map}) --}
(a) $\rm X^0$  position versus $\rm X^0$ FWHM. The black filled squares show
the mean values with standard deviations as error bars (b) Total PL intensity versus ratio of $\rm X^0$ intensity to total PL intensity, in which squares and triangles represent strongly and weakly coupled points respectively. The data from only one of the two MoSe$_2$/3L graphene regions in Sample 1 is given here for clarity.} 
    \label{Fig_S3_Correl}
\end{figure}

\begin{figure}[htb!]
    \begin{center}
    \includegraphics[width=1\linewidth]{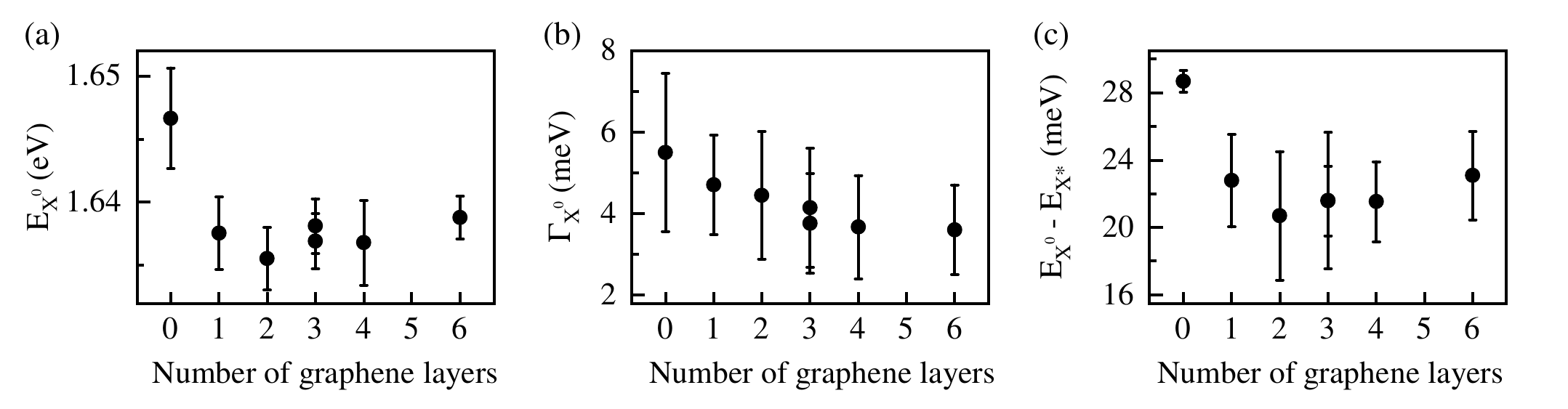}
    \caption{\textbf{Spatially averaged values from the PL mapping of Sample 1 --} (a,b) Peak energy and FWHM of the Bright exciton ($\rm X^0$) PL feature and (c) Energy difference between bright exciton and trion energy, i.e., trion binding energy extracted from The error bars correspond to the standard deviations around the spatially averaged values.}
    \label{fig_SI_S3_mean}
    \end{center}
\end{figure}

Figure~\ref{FigSM_S3Id}c shows representative PL spectra of each domain of Sample 1, recorded at ($T\sim$ 16 K). The PL spectrum of bare MoSe$_2$ is dominated by two narrow lines of a few meV in full-width at half maximum (FWHM)~(Fig.~\ref{fig_SI_S3_Map} and Fig.~~\ref{Fig_S3_Correl}) arising from bright neutral excitons ($\rm X^0$) and charged excitons (trions, $\rm X^{\star}$). As previously reported, the most striking effect of the graphene layers is to ``filter'' the MoSe$_2$ emission, leaving one single emission line arising from $\rm X^0$~\cite{Lorchat2020,Parralopez2021}. In particular, the spectral weight from charged excitons (trions, $\rm X^{\star}$) in MoSe$_2/N$-layer graphene ($N$LG) is one order of magnitude weaker than in the MoSe$_2$ reference, but still remains measurable. Due to dielectric screening~\cite{Raja2017,Lorchat2020}, the $\rm X^0$ line is slightly redshifted by about 10~meV in MoSe$_2$/$N$LG with fluctuations in the peak position and linewidth that we attribute to local variations of built-in strain, electrostatic environment and dielectric screening~\cite{Raja2019}. The spatially averaged shift~(Fig.~\ref{fig_SI_S3_mean}) does not significantly depend on the number of graphene layers suggesting that dielectric screening is mostly determined by the first coupled atomic layer~\cite{Moczko2025}. The intensity of the $\rm X^0$ PL line is quenched by more than one order of magnitude in MoSe$_2/N$LG with respect to the MoSe$_2$ reference. Interestingly, this quenching factor only increases weakly but still appreciably with increasing $N$ (Fig.~3 and Fig.~\ref{fig_SI_S3_mean}). The PL quenching effect and the filtering effect are consistently found in the large maps $(50\times120\: \mu\rm m^2)$ plotted in Fig.~\ref{FigSM_S3Id}d,e, showing the total PL intensity defined here as the sum of the $\rm X^0$ and $\rm X^{\star}$ PL intensities and the integrated intensity ratio of the line $\rm X^0$, $I_{\mathrm X^0}$ with respect to the total PL intensity $I_{\mathrm {tot}}$, respectively. 

To get further insights into the maps plotted in Fig.~\ref{FigSM_S3Id} and \ref{fig_SI_S3_Map} and quantitatively assess the sample homogeneity, we plot the correlation between the FWHM of the $\rm X^0$ line $\Gamma_{\mathrm X^0}$ and its energy $E_{\mathrm X^0}$ (Fig.~\ref{Fig_S3_Correl}a) for all spots fitted on Fig.~~\ref{FigSM_S3Id} and \ref{fig_SI_S3_Map}. The spatially averaged values over domains that largely exceed $100\:\mu \mathrm m^2$ are shown in Fig.~\ref{fig_SI_S3_mean} along with the associated standard deviations.

\clearpage

 Fig.~\ref{Fig_S3_Correl}a reveals dense clouds of points that confirm the qualitative trends identified in Fig.~~\ref{FigSM_S3Id} and \ref{fig_SI_S3_Map}. The spatially averaged FWHM of the $\rm X_0$ range from $5.5\pm2.0$~meV in hBN supported MoSe$_2$ down to $3.6\pm1.1$~\rm meV in hBN/MoSe$_2$/6LG. The FWHM values are quite typical from hBN- and/or graphene-capped samples and entail here a small contribution on the order of 1-2 meV due to the spectral and spatial resolutions of our setup. We note that the monotonic decrease in $\Gamma_{\mathrm X^0}$ as $N$ increases suggests a more homogeneous encapsulation of MoSe$_2$ by thicker graphene flakes. The peak positions are also narrowly distributed, typically with standard deviations of 3 to 6~meV depending on the region of interest. This dispersion arises from local changes in built-in strain and dielectric screening~\cite{Raja2019}.

The efficiency and spatial homogeneity of the filtering effect introduced above is analyzed through the correlation between the ratio $I_{\mathrm X^0}/I_{\mathrm {tot}}$ and $I_{\mathrm {tot}}$ shown in Fig.~\ref{Fig_S3_Correl}b. We observe a clear correlation between low $I_{\mathrm{tot}}$ (i.e., strong PL quenching) and weak trion intensity $I_{\mathrm X^{\star}}$ (efficient filtering).

\subsubsection{\textbf{TRPL measurements on Sample 1}}

\begin{figure}[ht!]
    \centering
    \includegraphics[width=1\linewidth]{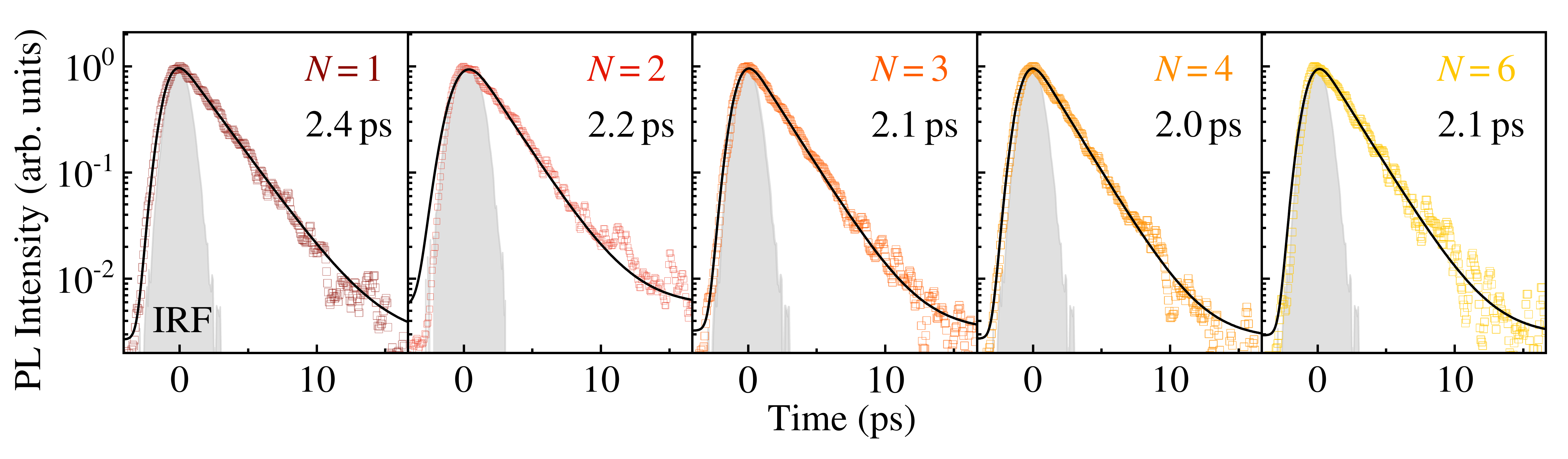}
    \caption{\textbf{TRPL measurements on Sample 1 --} TRPL measurements in all the MoSe$_2/N$-layer graphene domains investigated in Sample 1 (see Fig.~2d for selected data). The solid lines are fits to the data (symbols) using the three level system described in the main text and in Sec. \ref{SecRate}. The $\rm X^0$ decay times are indicated. All data were recorded at 6~K under laser excitation at 1.71~eV with a temporal resolution of 1~ps. The instrument response function (IRF) of the streak camera is shown in gray in each plot.} 
    \label{Fig_S3_TRPL_all}
\end{figure}

\clearpage

\subsubsection{\textbf{Photoluminescence excitation spectroscopy (PLE) on Sample 1}}

\begin{figure}[ht!]
    \centering
    \includegraphics[width=0.62\linewidth]{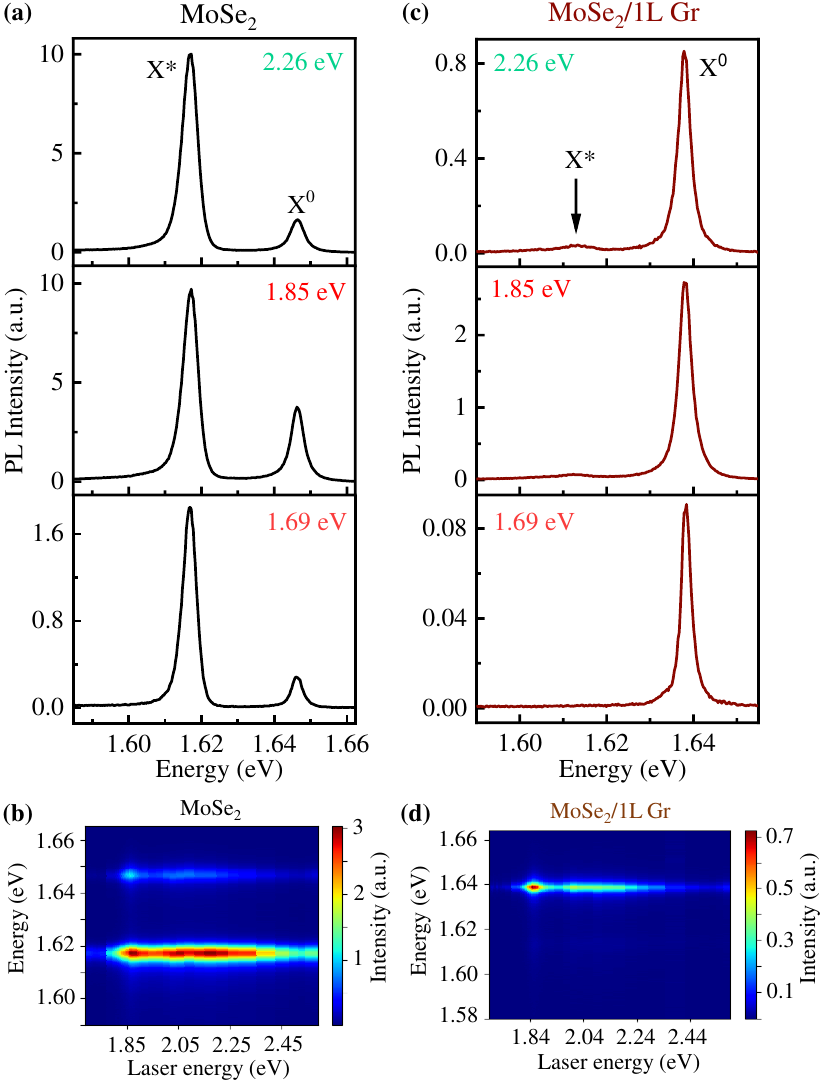}
    \caption{\textbf{Selected PLE spectra on Sample 1 --}
PL spectra of MoSe$_2$ recorded at three different laser photon energies  are shown in (a) and the full PLE scan shown as a colormap in (b). (c) and (d) show the results for the MoSe$_2$/1LG domain of Sample 1.  All measurements were performed at 16~K with a similar laser intensity close to 10 $\mu$W/$\mu$m$^2$. The PL intensities at each laser photon energy are normalized by the laser intensity.} 
    \label{Fig_S3_PLE_spectra}
\end{figure}

\begin{figure}[ht!]
    \centering
    \includegraphics[width=0.48\linewidth]{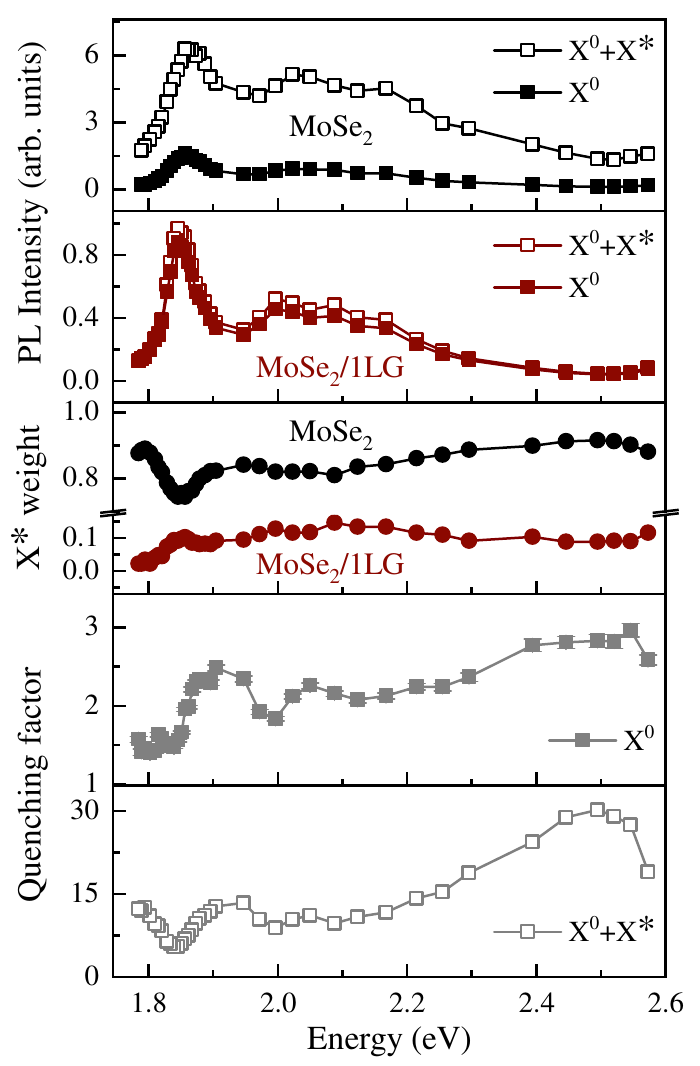}
    \caption{\textbf{PLE on Sample 1 --}
The panels from top to bottom showing PLE results: the $\rm X^0$ intensity and total PL
intensity for MoSe$_2$, the $\rm X^0$ intensity and total PL intensity for MoSe$_2$/1L, weight of $\rm X^{\star}$ in the PL intensity for MoSe$_2$ and MoSe$_2$/1L graphene, the quenching factors of $\rm X^0$ and total PL as a function of incident laser energy. The dotted and dashed lines show positions of $\rm X^0 _{2s}$ and $\rm B_{1s}$ exciton respectively. All measurements were performed at 16~K with a similar laser intensity close to 10 $\mu$W/$\mu$m$^2$. The PL intensities at each laser photon energy are normalized by the laser intensity.} 
    \label{Fig_S3_PLE}
\end{figure}

In order to investigate the influence of the laser photon energy on the near-field transfer between MoSe$_2$ and graphene, we have performed photoluminescence excitation spectroscopy (PLE) on the MoSe$_2$ reference and on the MoSe$_2$/1LG domain in Sample 1. In this experiment, the PL spectra are recorded in the linear excitation regime as a function of the incoming photon energy. Selected spectra are plotted in Fig.~\ref{Fig_S3_PLE_spectra} and the relevant parameters associated with the full PLE run are displayed in Fig.~\ref{Fig_S3_PLE}. At first glance, the PLE data reflects the optical response of MoSe$_2$ with a prominent resonance near the $B$ exciton~\cite{Wang2018}. The quenching factor for $\mathrm X^0$ emission drops appreciably for optical excitation below the B exciton and tends to increase again slightly as the laser excitation approaches the $\mathrm{X^0_{1s}}$ energy. This trend is clearer if one considers the ``total'' quenching factor. \\

The faint, yet measurable, spectral weight of $\mathrm X^{\star}$ emission in MoSe$_2$/1LG drops from typically $\lesssim 15\%$ around and above the $B$ exciton down to hardly measurable values lower than $1\%$ below the $B$ as the photon energy, illustrating a perfect ``filtering effect'' as the laser photon energy approaches the $\mathrm X_0$ line. Conversely, the spectral weight of $\mathrm X^{\star}$ emission remains large, typically above $80~\%$ in MoSe$_2$ and weakly dependent on the laser photon energy, although we notice a drop in the $\mathrm X^{\star}$ spectral weight down to $75~\%$ upon resonant excitation of the $B$ exciton followed by a continuous increase slightly above $90~\%$ upon non-resonant excitation below the $\mathrm {X^0_{2s}}$ exciton. These results suggest that residual $\mathrm X^{\star}$ formation in MoSe$_2$/1LG may mostly arise from  the photogenerated hot electrons and holes.
Interestingly, we notice that the $\mathrm X^{\star}$ and $\mathrm X^0$  FWHM are comparable in MoSe$_2$ and are in the range $4-5~\mathrm meV$, whereas in MoSe$_2$/1LG, the $\mathrm X^{\star}$ FWHM (10-20~meV) is about five times broader than the $\mathrm X^{\star}$ FWHM (2.5-3.5~meV). This result suggests that the residual $\mathrm X^{\star}$ emission may also have an extrinsic origin and reflect the nanoscale inhomogeneities in our sample.

\paragraph*{\textbf{Discussion on the PL FWHM} --}

Generally, we observe that the $\mathrm X^0$ FWHM tends to be slightly smaller in MoSe$_2$/1LG compared to bare MoSe$_2$, which may be in part due to the fact that the MoSe$_2$ reference  rests on SiO$_2$ in the case of Samples 1 and 4. 
At this stage and in contrast to previous reports~\cite{Hill2017,Tebbe2024}, we have never noticed a significant broadening in MoSe$_2-N$LG that we could assign to lifetime-induced broadening due to picosecond charge and/or energy transfer. 

\begin{figure}[ht!]
    \centering
    \includegraphics[width=0.5\linewidth]{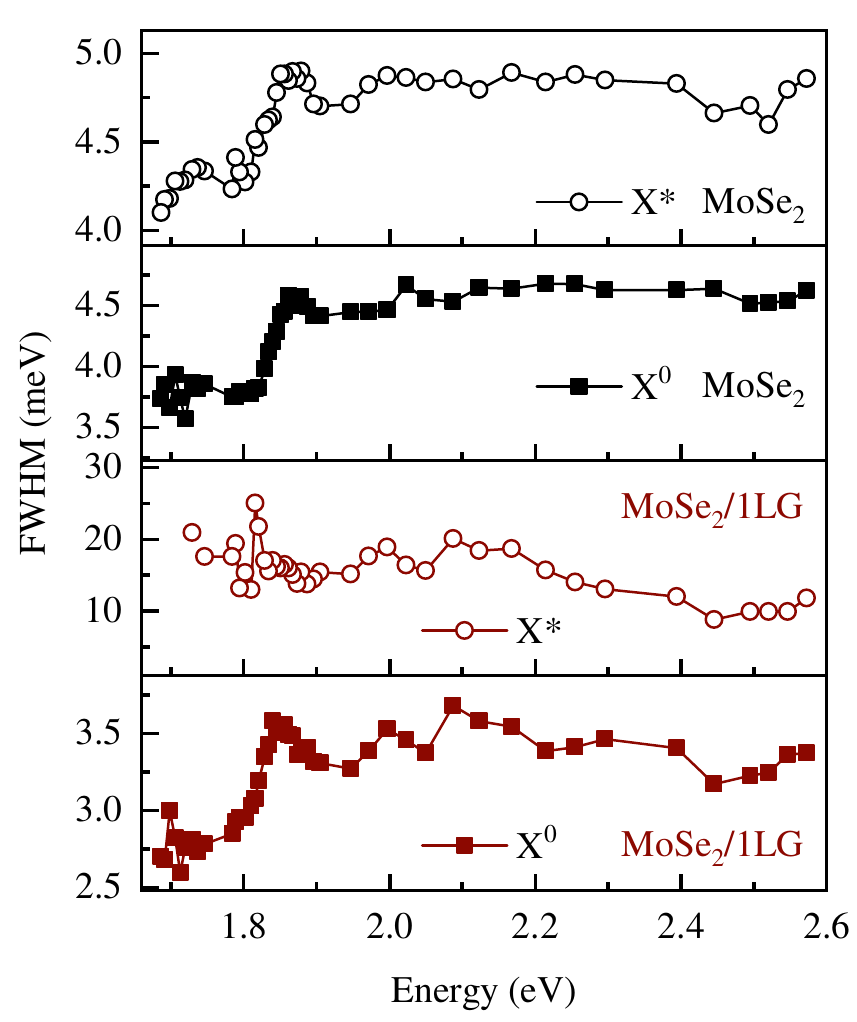}
    \caption{\textbf{PLE on Sample 1: FWHM --}
The panels from top to bottom showing FWHM of  PLE results: the $\rm X^0$ and  $\rm X^{\star}$  features MoSe$_2$ and MoSe$_2$/1LG, respectively, as a function of incident laser photon energy. All measurements were performed at 16 K with a similar laser intensity close to 10 $\mu$W/$\mu$m$^2$.} 
    \label{Fig_S3_PLE_FWHM}
\end{figure}

Noteworthy, the $\mathrm X^0$ FWHM of MoSe$_2$/1LG  MoSe$_2$ as well as the $\rm X^{\star}$ FWHM of MoSe$_2$  narrow appreciably by about 1~meV for excitation below the $B$ exciton (Fig.~\ref{Fig_S3_PLE_FWHM}). This narrowing may be due do the lower exciton density achieved under non-resonant excitation. We also note that the FWHM of the faint residual $\rm X^{\star}$ emission remains much larger in MoSe$_2$/1LG, typically above 10~meV. This points towards an extrinsic origin of $\rm X^{\star}$ PL in MoSe$_2$/1LG that tends to vanish as the laser photon energy approaches the $\rm X^0$ resonance.

\clearpage

\subsubsection{\textbf{Differential reflectance spectroscopy on Sample 1}}

\begin{figure}[htb!]
    \begin{center}
    \includegraphics[width=0.33\linewidth]{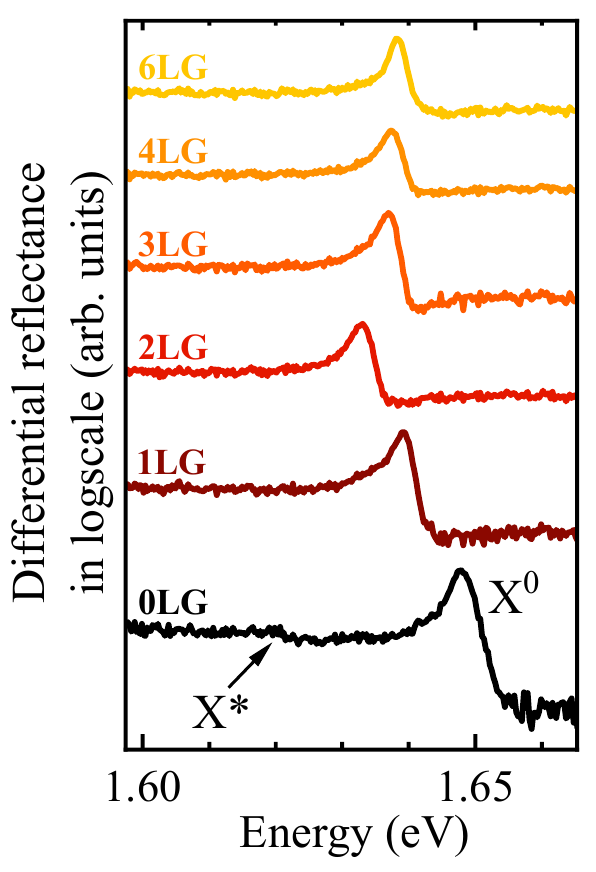}
    \caption{\textbf{Differential reflectance (DR/R) spectra from Sample 1 --} The spectra (shown here in log scale and vertically shifted for clarity) were recorded at cryogenic temperature (16~K) under white light illumination for the different domains identified in Sample 1 (Fig.~\ref{fig_SM_GrThickness} and Fig.~\ref{FigSM_S3Id}). Besides the main excitonic feature $\rm X^0$, a faint feature arising from trion ($\rm X^{\star}$) absorption is observable on the MoSe$_2$ monolayer reference.}
    \label{fig_fit}
    \end{center}
\end{figure}

\clearpage

\subsection{\textbf{Sample 2}}
\label{SecSMSample2}
\begin{figure}[ht!]
    \centering
    \includegraphics[width=1\linewidth]{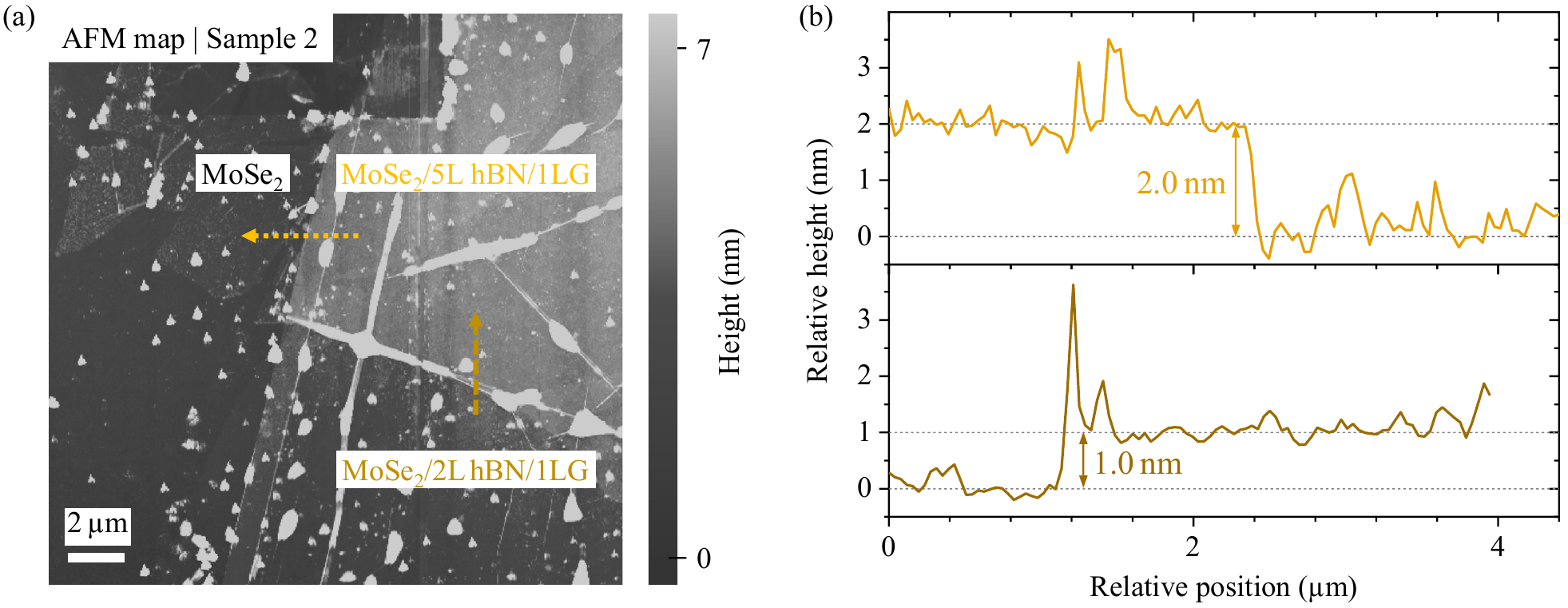}
    \caption{\textbf{AFM measurements on Sample 2  --}. (a) AFM  map of Sample 2  and (b) selected height profiles indicated with dashed arrows in (a).} 
    \label{Fig_S2_AFM}
\end{figure}

An atomic force microscopy (AFM) map was measured on Sample 2 (see the sequence of its layers in Fig.~\ref{Fig_SM_samples}) in the vicinity of the MoSe$_2/N$-layer hBN/1-layer graphene region, in order to quantitatively assess the thickness of the hBN spacer (see Fig.~\ref{Fig_S2_AFM}). Although the AFM map was recorded on the full stack, including the top hBN layer, one can clearly resolve discrete steps of 2.0~nm and 1.0~nm, respectively, between MoSe$_2$ and two MoSe$_2/N$-layer hBN/1-layer graphene domains. Considering the nearly identical nominal layer thicknesses of hBN and graphene (near 0.33~nm) we estimate that the hBN spacer displays two domains with thicknesses of 2 and 5 atomic layers, respectively.

\clearpage

\subsection{Sample 3}
\label{SecSMSample3}
An AFM map was measured on Sample 3 (see the sequence of its layers in Fig.~\ref{Fig_SM_samples}) in the vicinity of the MoSe$_2/N$-layer hBN/1-layer graphene region, in order to quantitatively assess the thickness of the hBN spacer. The map was recorded on the fully stack, including the top hBN layer. One can clearly resolve a discrete step of approximately 5~nm between MoSe$_2$/1-layer graphene and MoSe$_2/N$-layer hBN/1-layer graphene domains. We therefore estimate that the hBN spacer is approximately 15 atomic layers thick.

\begin{figure}[ht!]
    \centering
    \includegraphics[width=1\linewidth]{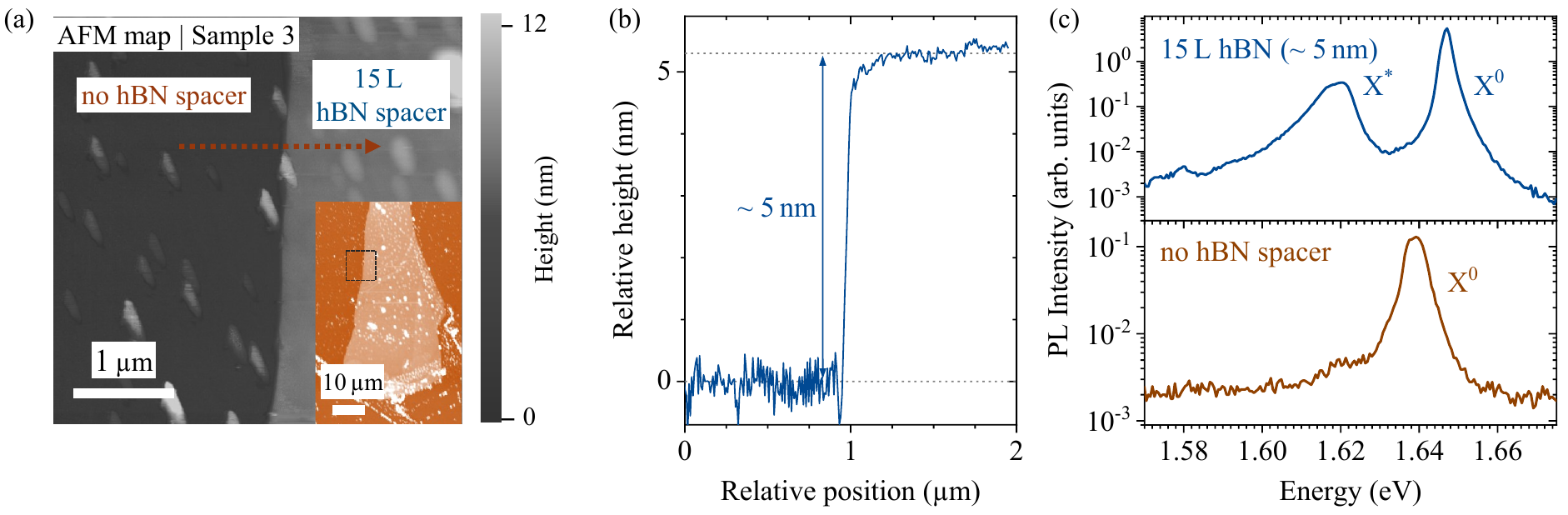}
    \caption{\textbf{AFM measurements and PL spectra of Sample 3  --} (a) AFM  map of Sample 3. The inset shows an AFM map over a larger area of the sample, where the hBN spacer flake is clearly visible. (b) Selected height profile indicated with a dashed arrow in (a). (c) PL spectra of the MoSe$_2$/15L\:hBN/Gr (top, blue) and MoSe$_2$/Gr regions (bottom, brown).} 
    \label{Fig_S3_AFM_PL}
\end{figure}

Low temperature (4~K) PL spectra of the MoSe$_2$/15L\:hBN/Gr and MoSe$_2$/Gr regions are shown in  Figure~\ref{Fig_S3_AFM_PL}c. The  MoSe$_2$/Gr PL spectrum is very similar to the measurements on Samples 1 (Fig.~1 and \ref{FigSM_S3Id}) and 5 (Fig.~\ref{Fig_Luis_PL_TRPL}a), whereas the MoSe$_2$/15L\:hBN/Gr PL spectrum resembles that of an MoSe$_2$ reference with minimal residual doping (akin to the reference in Sample 5, see Fig.~\ref{Fig_Luis_PL_TRPL}a). The corresponding TRPL measurements on Sample 3 are shown in Fig.~3a of the main manuscript.

\clearpage

\subsection{\textbf{Sample 4}}
\label{SecSMSample4}
\begin{figure}[ht!]
    \centering
    \includegraphics[width=1\linewidth]{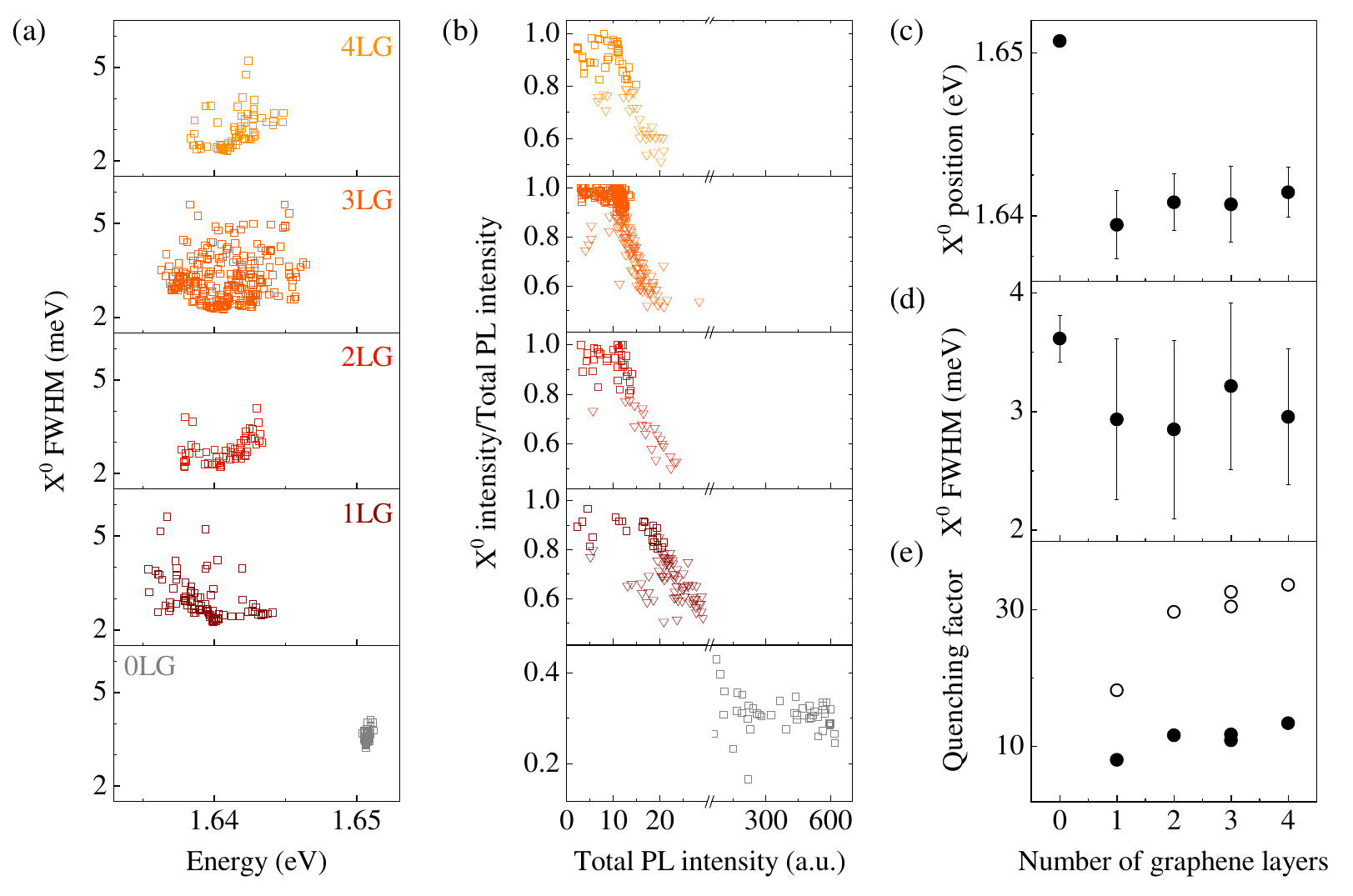}
    \caption{\textbf{ Correlation plots from PL mapping of Sample 4  --} (a) Correlation between $\rm X^0$  position and $\rm X^0$ FWHM, the black filled squares show the mean values with standard deviations as error bars (b) Correlation between total PL intensity and ratio of $\rm X^0$ intensity to total PL intensity. All correlation plots are obtained from maps recorded at a temperature of 16~K, under laser illumination at 1.96~eV. (c-e) mean values of (c) $\rm X^0$ position, (d) $\rm X^0$ FWHM and (e) quenching factors as a function of number of graphene layers under MoSe$_2$ taken from PL maps. The open and filled symbols in (e) represent quenching factors for total PL and for $\rm X^0$, respectively. } 
    \label{Fig_S8_Mapping}
\end{figure}

\begin{figure}[ht!]
    \centering
    \includegraphics[width=0.55\linewidth]{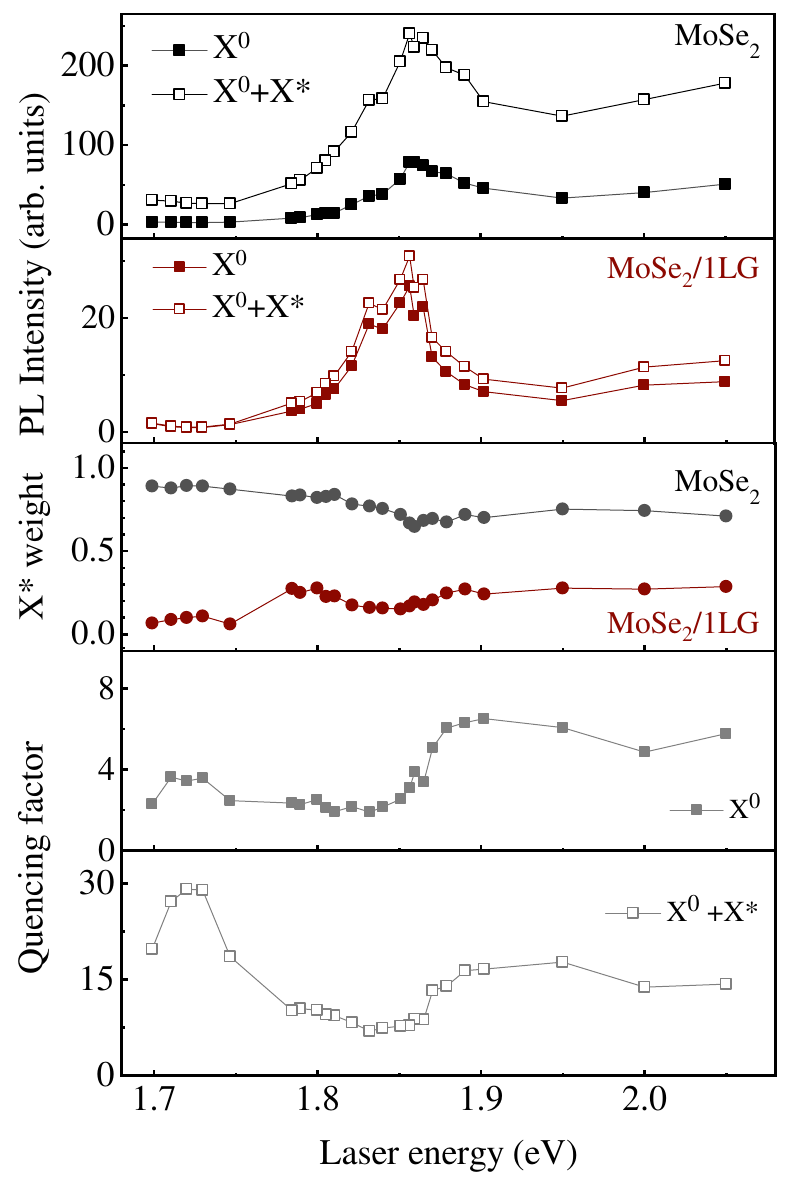}
    \caption{\textbf{PLE on Sample 4 --}
PLE measurements for Sample 4, similar to Sample 1, displayed as in Fig.~\ref{Fig_S3_PLE}. Measurements were performed at 16 K with a laser intensity close to 10 $\mu$W/$\mu$m$^2$.} 
    \label{Fig_S8_PLE}
\end{figure}

Figures~\ref{Fig_S8_Mapping} and ~\ref{Fig_S8_PLE} display PL data for Sample 4, another sample akin to Sample 1 (see Fig.~\ref{Fig_SM_samples}) containing an hBN-capped MoSe$_2$ reference on SiO$_2$ as well as hBN-capped MoSe$_2-N$LG, with $N=1,2,3,4$, in particular with the same underlying SiO$_2$ epilayer thickness of 500~nm.

\clearpage

\subsection{\textbf{Sample 5}}
\label{SecSMSample5}
\begin{figure}[ht!]
    \centering
    \includegraphics[width=1.\linewidth]{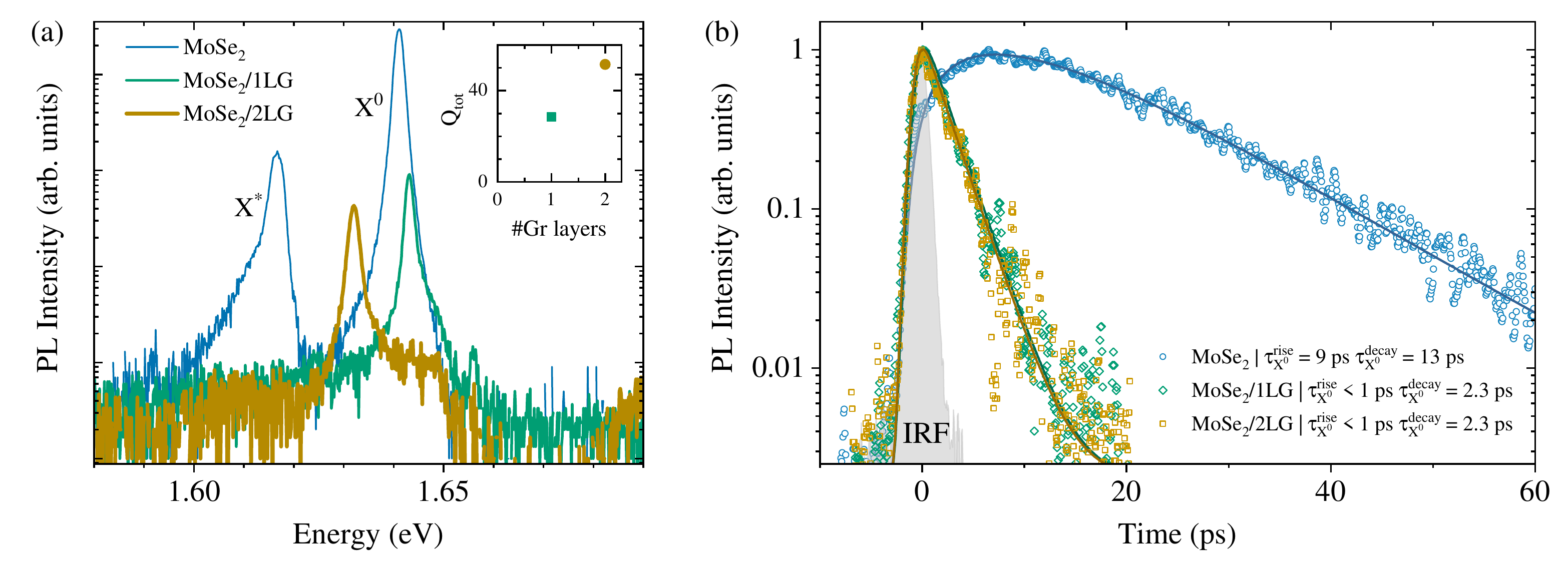}
    \caption{\textbf{PL and TRPL on Sample 5 --}
 (a) PL spectra of hBN-encapsulated MoSe$_2$, MoSe$_2$/1LG and  MoSe$_2$/2LG in Sample 5, recorded in the linear excitation regime with PL intensity shown in log-scale. The total PL quenching factors $Q_{\mathrm {tot}}$ are shown in an inset. (b) TRPL of the hBN-encapsulated MoSe$_2$ reference, MoSe$_2$/1LG and MoSe$_2$/2LG. All data were recorded under laser excitation at 1.71~eV with a temporal resolution of 1~ps. The instrument response function (IRF) of the streak camera is shown in gray in (b).} 
    \label{Fig_Luis_PL_TRPL}
\end{figure}

Sample 5 is an hBN-encapsulated MoSe$_2$/graphene sample made with a relatively thick bottom hBN layer of about 120~nm  and a 90-nm thick SiO$_2$ layer (see Fig.~\ref{Fig_SM_samples}). In these conditions, MoSe$_2$ lies at a node of the electromagnetic field and a long $\rm X^0$ radiative lifetime is expected and experimentally observed as shown in Fig.~\ref{Fig_Luis_PL_TRPL} below~\cite{Fang2019}. The graphene flake has a monolayer (1LG) and a bilayer (2LG) domain and in Fig.~\ref{Fig_Luis_PL_TRPL} below, we discuss PL and TRPL measurements on the hBN-encapsulated MoSe$_2$ reference, MoSe$_2$/1LG and MoSe$_2$/2LG.

Noteworthy, in contrast with Samples 1 and 4, the MoSe$_2$ reference in Sample 5 shows much fainter trion ($\rm X^{\star}$) emission, more than one order of magnitude lower than $\mathrm {X^0}$ emission, suggesting minimal unintentional doping (Fig.~\ref{Fig_Luis_PL_TRPL}a). The origins of these differences may arise from the bulk MoSe$_2$ flakes as well as from the local electrostatic environment. 

The $\rm X^0$ TRPL of the MoSe$_2$ reference shows a well-resolved rise time of  $(9\pm1)\:\rm {ps}$ and a decay time of $(13\pm 1)\:\mathrm{ps}$, which, following Ref.~\cite{Fang2019} we assign to $\rm X^0$ decay and hot exciton relaxation, respectively. The TRPL measurements on MoSe$_2$/1LG and  MoSe$_2$/2LG regions both reveal the same decay time of $(2.3\pm 0.1)\:\mathrm{ps}$ assigned to the $\rm X^0$ lifetime, and an unresolved rise arising from hot exciton relaxation. The measured $\rm X^0$ lifetime is very similar to the value measured in Samples 1 and 3.

We note that unlike in Sample 1, we are able to resolve a PL rise and decay time in the MoSe$_2$ reference of Sample 5, whereas the PL rise time in the MoSe$_2$ reference of Sample 1 was not resolved. We therefore suggest that native doping has a strong impact on hot exciton relaxation and leads to the observation of much shorter PL rise times in the $\rm X^0$ TRPL scans. Such extrinsic effects will be addressed in subsequent studies.

In closing, we stress that in spite of the very different emission properties of the MoSe$_2$ references in our four samples, we consistently see very similar PL quenching factors and exciton dynamics in MoSe$_2$/graphene. 



\clearpage

\section{\textbf{FRET from MoSe$_2$ excitons to multilayer graphene}}
\label{SecSotos}
  In this section, we model the F\"orster-type resonant energy transfer (FRET) between MoSe$_2$ excitons and adjacent graphene layers using a classical electrodynamic approach. Specifically, we study how the local density of optical states (LDOS) at the position of the exciton is modified as a function of the number of graphene layers. The system assumes an electric dipole in a stratified dielectric environment and by using generalized Fresnel coefficients and dielectric properties of the materials involved, we compute the dissipated power spectrum of the exciton. 

\subsection{\textbf{Fresnel coefficients of multilayer structures}}

To model the radiative and non-radiative decay of point-like electric dipoles inside a multilayer heterostructure we begin by formulating the Fresnel reflection coefficients.

\begin{figure}[htb!]
    \centering
    \includegraphics[width=1\linewidth]{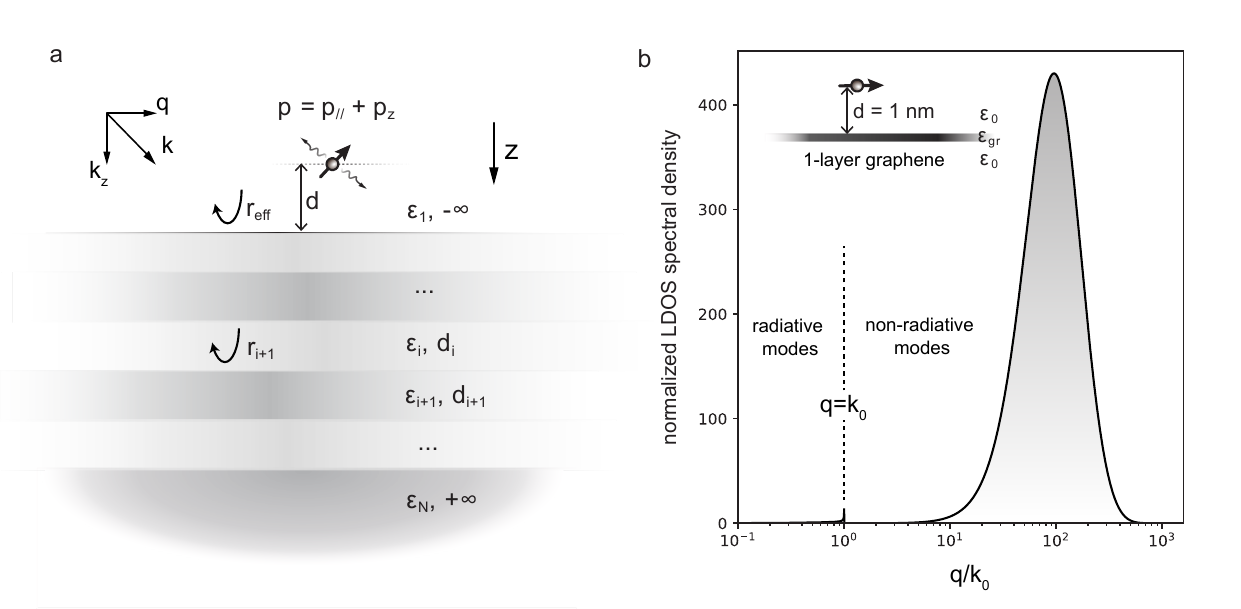}
    \caption{(a) Schematic representation of the dipole emitter (with both in-plane and out-of-plane components) placed at distance $d$ above a multilayer stack characterized by dielectric functions $\varepsilon_i$ and thicknesses $ d_i$. The structure terminates in semi-infinite media above and below. (b) Normalized angular spectrum of the LDOS of an in-plane dipole near a single-layer graphene at $d = 1$~nm, plotted as a function of in-plane wavevector $ \qpar/k_0 $, showing contributions from radiative and non-radiative modes. A strong enhancement is seen in the high-$ \qpar$ (non-radiative modes) which is attributed to FRET interactions.}

    \label{fig1}
\end{figure}

\paragraph*{\textbf{Iterative formulation}}

Consider a system composed of $N$ dielectric layers stacked along the $z$-axis, each characterized by a permittivity $\varepsilon_i$, thickness $d_i$ and bounded by semi-infinite media above and below as shown in Fig.~\ref{fig1}a. The wavevector component parallel to the plane of the layers is $\mathbf{q}$ and the out-of-plane wavevector component (z-component) in layer $i$ is given by

\begin{equation}
    k_{z,i} = k_i^2 - \qpar^2.
\end{equation}


where $k_i = \sqrt{\varepsilon_i} \left( \frac{\omega}{c} \right)^2 = \sqrt{\varepsilon_i}  k_0 ^2$ is the absolute value of the wave vector in layer $i$. For a dipole located above the multilayer stack at distance $d$, the total reflection coefficients for $p$- and $s$-polarized waves (TM and TE, respectively) can be computed iteratively using the generalized Fresnel relations. Starting from the last interface (between layer $N-1$ and layer $N$). The effective reflection coefficients are calculated using \cite{Chew1995} :

\begin{equation}
    r_{p,s}^{(i)} = \frac{r_{p,s}^{(i-1,i)} + r_{p,s}^{(i+1)} e^{2i k_{z,i+1} d_{i+1}}}{1 + r_{p,s}^{(i-1,i)} r_{p,s}^{(i+1)} e^{2i k_{z,i+1} d_{i+1}}},
\end{equation}

where $r_{p,s}^{(i-1,i)}$ are the standard Fresnel coefficients at the interface between layers $i-1$ and $i$, given by:

\paragraph{For $s$-polarization (TE):}
\begin{equation}
    r_s^{(i-1,i)} = \frac{k_{z,i-1} - k_{z,i}}{k_{z,i-1} + k_{z,i}},
\end{equation}

\paragraph{For $p$-polarization (TM):}
\begin{equation}
    r_p^{(i-1,i)} = \frac{\varepsilon_{i} k_{z,i-1} - \varepsilon_{i-1} k_{z,i}}{\varepsilon_{i} k_{z,i-1} + \varepsilon_{i-1} k_{z,i}}.
\end{equation}

Note that $2<i<N$ and in the case of $i+1>N$ then $r_{p,s}^{(i+1)}=0$. The total effective reflection coefficient is then given by  $r_{p,s}^{\rm{eff}}=r_{p,s}^{(2)}$. Due to the dipole distance $d$ of the dipole from the structure an associated phase shift of $e^{2i k_{z,1} d}$ needs to be taken into account. These coefficients form the basis for calculating how the presence of the multilayer structure modifies the dipole's radiative and non-radiative coupling by helping us calculate the dissipated power of the dipole. 

\subsection{\textbf{Dissipated power of a dipole next to a multilayer structure}}

The total power dissipated by a classical dipole near a multilayer interface can be computed by evaluating the work done by the reflected field back on the dipole. The time-averaged power dissipated by a dipole with moment $\mathbf{p}$ located at position $\mathbf{r}_d$ is given by \cite{Novotny2012},

\begin{equation}
    \mathcal{P} = \frac{\omega}{2} \, \mathrm{Im} \left[ \mathbf{p}^{\star} \cdot \mathbf{E}_{\text{tot}}(\mathbf{r}_d) \right],
\end{equation}

where $\mathbf{E}_{\text{tot}}$ is the total electric field at the location of the dipole, including contributions from the emitted field and the field reflected by the multilayer structure and $[\dots]^{\star}$ denotes the complex conjugate. The electric field at the dipole position is expressed using the generalized Fresnel coefficients $r_{p,s}^{\text{eff}}$ derived previously. The reflected field includes the information of the electromagnetic environment due to the presence of the stack and can be computed by integrating over $\qpar$. Then the dissipated power is given by~\cite{Ford1984}

\begin{multline}
\mathcal{P} = \frac{\omega}{2 \varepsilon_1} \, \mathrm{Re} \int_0^{\infty} \, \frac{\qpar}{k_{z,1}}  \left[ |\mathbf{p}_\perp|^2 \left( 1 + r_p^{\text{eff}} e^{2i k_{z,1} d} \right) + \frac{1}{2} k_{1}^2 |\mathbf{p}_\parallel|^2 \left( 1 + r_s^{\text{eff}} e^{2i k_{z,1} d} \right) \right. \\ \left. + \frac{1}{2} k_{z,1}^2 |\mathbf{p}_\parallel|^2 \left( 1 - r_p^{\text{eff}} e^{2i k_{z,1} d} \right) \right]  \mathrm{d}\qpar.
\end{multline}

Here, $\mathbf{p}_\perp$ and $\mathbf{p}_\parallel$ are the dipole moment components perpendicular and parallel to the multilayer interface, respectively. 

This expression comes from the angular spectrum representation of the dipole field~\cite{Ford1984}, where the total field is decomposed into plane-wave components with wavevector $\qpar$. The angular spectrum formulation shows how each $\qpar$ component of the dipole field interacts with the multilayer system through the Fresnel coefficients $r_p^{\text{eff}}$ and $r_s^{\text{eff}}$. The exponential terms express the round-trip phase delay between the dipole and the structure as commented before. 

The separation between radiative and non-radiative decay channels is possible through the integrand in the expression for $\mathcal{P}$ and can be interpreted as the power spectral density per in-plane wavevector $\qpar$, which reflects how much each spatial frequency (momentum) contributes to the total dissipation. Specifically, the quantity

\begin{equation}
    \frac{\mathrm{d}\mathcal{P}}{\mathrm{d}\qpar} \cdot \frac{\qpar}{\mathcal{P}_\text{ref}}
\end{equation}

can be plotted to visualize the angular spectrum, normalized to the power radiated in a homogeneous reference medium $\mathcal{P}_\text{ref}$, usually vacuum. As an example, we plot in Fig.~\ref{fig1}b the angular spectrum of the dissipated power for a dipole placed 1 nm above a single graphene layer in vacuum. This expresses the normalized LDOS spectral density. 

The angular spectrum presents two distinct regions: modes with $\qpar < k_0$, which correspond to radiative plane waves propagating into the far field (radiative modes) and modes with $\qpar > k_0$, which are evanescent and decay exponentially away from the dipole (non-radiative modes). These high-$\qpar$ modes are responsible for near-field coupling to lossy channels such as graphene and thus dominate the non-radiative energy transfer regime relevant for FRET.

\begin{figure}[ht!]
    \centering
    \includegraphics[width=1.0\linewidth]{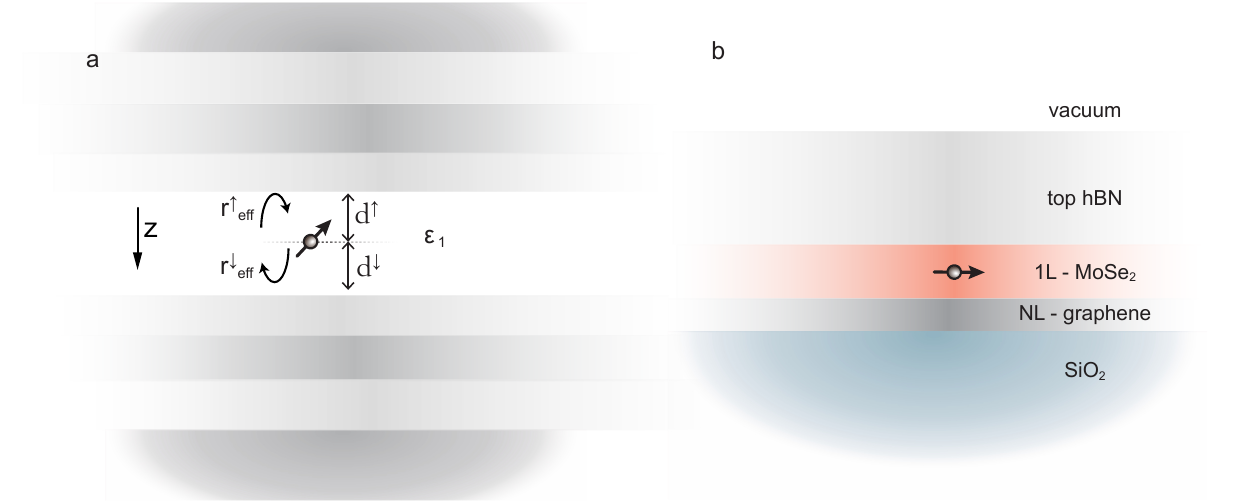}
    \caption{(a) Schematic depiction of the upward and downward reflection coefficients ($ r^\uparrow_{\text{eff}} $, $ r^\downarrow_{\text{eff}} $) used in the calculation of Eq. 10 when the dipole is embedded within a layer with $ \varepsilon = \varepsilon_1$. (b) The layered heterostructure considered in the LDOS calculation. The structure consists of a monolayer MoSe$_2$ placed between top hBN and multilayer graphene, resting on a SiO$_2$ substrate. The dipole emitter is located within the MoSe$_2$ layer. }

    \label{fig2}
\end{figure}

The local density of optical states (LDOS) enhancement at the dipole's position is given by the ratio
\begin{equation}
    \frac{\rho(\mathbf{r}_d)}{\rho_0} = \frac{\mathcal{P}(\mathbf{r}_d)}{\mathcal{P}_0},
\end{equation}

where $\rho_0$ and $\mathcal{P}_0$ are the local density of states and dissipated power, respectively, for a dipole in a homogeneous vacuum environment.

Since the decay rate $\Gamma$ of the exciton is proportional to the LDOS, the normalized decay rate can be written as:

\begin{equation}
    \frac{\Gamma(\mathbf{r}_d)}{\Gamma_0} = \frac{\rho(\mathbf{r}_d)}{\rho_0}.
\end{equation}

where $\Gamma_0$ is the decay rate in vacuum.

To evaluate the dissipated power for a dipole located inside a specific layer of a multilayer structure as shown in Fig~\ref{fig1}, we first evaluate the total effective Fresnel reflection coefficients, as shown previously, for the structure over and below the dipole separately, and we use an extension of the angular spectrum formalism. Consider a dipole within a layer $i=1$ of thickness $L_1$ and dielectric constant $\varepsilon_1$ having a distance $d^{\downarrow}$ and $d^{\uparrow}$ from the first interface below and above it, respectively. The total power dissipated by this dipole can be expressed as \cite{Ford1984}:

\begin{equation}
\mathcal{P} = \frac{\omega}{2} \, \mathrm{Re} \int_0^\infty \, \frac{\qpar}{k_{z,1}} \left\{
|\mathbf{p}_\perp|^2 \cdot \frac{\qpar^2}{k_1^2} \cdot
\left[
\frac{\left(1 + r_p^{\downarrow} e^{2i k_{z,1} d^{\downarrow}} \right)
      \left(1 + r_p^{\uparrow} e^{-2i k_{z,1} d^{\uparrow}} \right)}%
     {1 - r_p^{\downarrow} r_p^{\uparrow} e^{-2i k_{z,d} L_1}}
\right]
\right.
\end{equation}

\begin{equation*}
\left.
+ \frac{1}{2} |\mathbf{p}_\parallel|^2 \cdot \left[
\frac{k_{z,1}^2}{k_1^2} \cdot
\frac{\left(1 + r_s^{\downarrow} e^{2i k_{z,1} d^{\downarrow}} \right)
      \left(1 + r_s^{\uparrow} e^{-2i k_{z,1} d^{\uparrow}} \right)}%
     {1 - r_s^{\downarrow} r_s^{\uparrow} e^{-2i k_{z,1} L_1}}
+
\frac{\qpar^2}{k_1^2} \cdot
\frac{\left(1 - r_p^{\downarrow} e^{2i k_{z,1} d^{\downarrow}} \right)
      \left(1 - r_p^{\uparrow} e^{-2i k_{z,1} d^{\uparrow}} \right)}%
     {1 - r_p^{\downarrow} r_p^{\uparrow} e^{-2i k_{z,1} L_1}}
\right]
\right\} \mathrm{d}\qpar 
\end{equation*}

The coefficients $r_s^{\uparrow}$ and $r_s^{\downarrow}$ represent the effective Fresnel reflection coefficients for the multilayer structure above and below the dipole respectively and are calculated with Eq. 2. The numerator terms account for the interference between waves reflected from the two interfaces, while the denominators include the Fabry-Pérot-like multiple reflections within the layer. 

With this formalism, we can compute the dissipated power for a dipole embedded within any layer of a multilayer structure. By specifying the position of the dipole, the thicknesses of the surrounding layers and the dielectric constants of each material, the model calculates the available decay channels. 

In our case, we apply this model to the specific layered structure shown in Fig.~\ref{fig2}b, which consists, from top to bottom, of: hexagonal boron nitride (hBN), a monolayer MoSe$_2$ flake (hosting the excitonic in-plane dipole) and a N-layer graphene and SiO$_2$. By varying the graphene layer number N, we examine how the electromagnetic environment and local density of optical states (LDOS) evolve. 

In this case we model the exciton as a classical dipole embedded in a lossless TMD layer, meaning that we keep only the real part of the dielectric constant of MoSe$_2$ in the calculations. This approximation is essential because the formalism used to compute the dissipated power assumes the dipole resides in a non-absorbing medium. Including an imaginary component in the permittivity at the dipole's position would lead to unphysically strong quenching.

This behavior can be understood intuitively by considering the classical dipole--dipole energy transfer rate $\kappa_\tau \propto 1/d^6$, where $d$ is the distance between the donor and acceptor dipoles~\cite{Novotny2012}. As the two dipoles approach each other $d \to 0$), the transfer rate diverges, highlighting the breakdown of the point-dipole approximation.

Physically, this approximation is justified by the fact that we are primarily interested in the electromagnetic environment that surrounds the exciton not its intrinsic material loss. Specifically, what governs the dipole's decay into non-radiative channels are the lossy materials of the surrounding environment and the material in which the dipole resides modifies this coupling only by dielectric screening. By using the real part of the TMD permittivity, we account for the dielectric screening of the field while excluding infinite dissipation that would incorrectly be attributed to the emitter medium itself.

This assumption is also adopted in other theoretical approaches modeling exciton-graphene~\cite{Selig2019}, where the TMD is treated as a purely dielectric environment that modifies the Coulomb potential.

Finally, it is important to note that the effective dielectric constant experienced by the dipole is not simply that of MoSe$_2$. Due to the finite thickness of the MoSe$_2$ layer, the dipole interacts with an electromagnetic environment that averages over surrounding materials, resulting in a reduced effective permittivity. This effect can be quantified by computing the LDOS enhancement in a structure without any graphene. In this case, we obtain a total dissipated power enhancement of $\mathcal{P} / \mathcal{P}_0 = 2.5$. 

The power radiated by a dipole in a homogeneous, non-absorbing dielectric with effective permittivity $\varepsilon_{\text{eff}}$ is given by
\begin{equation}
\mathcal{P}_{\text{diel}} = \frac{\sqrt{\varepsilon_{\text{eff}}} \, \omega^4 |\mathbf{p}|^2}{12\pi \varepsilon_0 c^3},
\end{equation}
while the corresponding power radiated in vacuum is
\begin{equation}
\mathcal{P}_0 = \frac{\omega^4 |\mathbf{p}|^2}{12\pi \varepsilon_0 c^3}.
\end{equation}
Taking the ratio yields
\begin{equation}
\frac{\mathcal{P}_{\text{diel}}}{\mathcal{P}_0} = \sqrt{\varepsilon_{\text{eff}}}.
\end{equation}
From the calculated LDOS enhancement $\mathcal{P}/\mathcal{P}_0 = 2.5$, we then infer
\begin{equation}
\varepsilon_{\text{eff}} \approx (2.5)^2 = 6.25.
\end{equation}

This effective dielectric constant will be used in the next section to estimate the relevant in-plane momentum range for the FRET interaction.

\subsection{\textbf{Dielectric modeling of graphene in a layered environment}}

To describe the dielectric properties of a monolayer graphene embedded within a van der Waals heterostructure, we first need to model its dielectric function as
\begin{equation}
\varepsilon_{\text{gr}}(\omega, k) = \varepsilon^{\text{gr}}_{\text{eff}} + \frac{i\sigma(\omega)}{\varepsilon_0 \omega d},
\end{equation}
where $\sigma(\omega)$ is the 2D optical conductivity of graphene, $d \approx 0.33$~nm is its effective thickness and $\varepsilon^{\text{gr}}_{\text{eff}}$  is the average background permittivity of the surrounding media. 
Following the analysis of \cite{Falkovsky2007}, in order to determine a simplified relation of graphene's optical conductivity we need to define the appropriate regime in terms of temperature $T$, photon (or exciton) energy $\hbar\omega$ and momentum $\qpar$ in our case $E = \hbar\omega = 1.55~eV$. The position of the maximum of the FRET-related angular spectrum of the dissipated power is a function of the distance of the dipole from graphene as well as its dielectric environment \cite{Ford1984}.

\begin{equation}
\qpar^{\text{peak}} \sim \frac{1}{d_{\text{dip}}\,\varepsilon_{\text{diel}}},
\end{equation}
where $d_{\text{dip}}$ is the emitter–graphene separation. In this case we need the effective dielectric constant seen by the dipole in the TMD $\varepsilon_{\text{eff}}$ as calculated in the previous section

Assuming $d_{\text{dip}} \approx 0.38$~nm and $\varepsilon_{\text{eff}} \approx 6.25$, we estimate

\begin{equation}
\hbar k^{peak}_\parallel v_F \approx \hbar \cdot \qpar \cdot v_F \approx 0.27\,\text{eV},
\end{equation}

with $v_F \approx 10^6$~m/s the graphene Fermi velocity.

Since  $\hbar \qpar v_F \approx 0.27\,\rm{eV} <  \hbar\omega = 1.55\,\rm{eV} $ , the graphene optical conductivity lies in the intermediate-wavevector interband regime. In this limit, it is well approximated by the asymptotic form derived by ~\cite{Falkovsky2007}:

\begin{equation}
\sigma(\omega) = \frac{e^2}{4\hbar} - \frac{2i e^2}{\pi \hbar} \cdot \frac{k_B T}{\omega} \left[ \ln 2 + 6 \zeta(3) \left( \frac{k_B T}{\hbar \omega} \right)^2 \right],
\end{equation}

where the first term corresponds to the universal interband conductivity of graphene and the second (imaginary) term represents a thermal correction. Here, $ \zeta(3) \approx 1.202 $ is the Riemann zeta function, which arises from the expansion of the Fermi–Dirac distribution at finite temperature.  In our case $T \approx 4~\rm K$ so $k_B T = 0.35~\rm{meV} \ll \hbar\omega = 1.55\,\rm{eV} $, which simplifies Eq. 13 to only the real part and we compute the effective complex dielectric constant of monolayer graphene as
\begin{equation*}
    \varepsilon_{\text{gr}}(\omega) \approx 8.75 + 8.84i,
\end{equation*}

where the real part reflects the averaged background permittivity due to the surrounding materials and the imaginary part captures the dominant absorptive response relevant for near-field energy transfer. In our structure, graphene is encapsulated between MoSe$_2$ and hBN layers and we take $\varepsilon_{\text{eff}}^{\text{gr}} \approx 8.75 $ as the arithmetic mean of their respective optical permittivities.

\subsection{\textbf{Application to MoSe$_2$/graphene structures}}

We now apply the formalism developed above to compute the LDOS enhancement experienced by an excitonic dipole embedded in a MoSe$_2$ monolayer in the presence of multilayer graphene. For multilayer graphene, we assume a simple additive model where each additional monolayer contributes linearly to the total surface conductivity \cite{Hongki2009}:

\begin{equation}
\sigma_N(\omega) = N \cdot \sigma(\omega),
\end{equation}
with $N$ the number of graphene layers (NL). In that way the corresponding multilayer dielectric function is estimated to be the same as in the case of a monolayer (1L) graphene as follows:
\begin{equation}
\varepsilon^{NL}_{\text{gr}}(\omega) =\varepsilon^{\text{gr}}_{\text{eff}}+ \frac{i \sigma_N(\omega)}{\varepsilon_0 \omega (N\cdot d)}=\varepsilon^{\text{gr}}_{\text{eff}}+ \frac{i  N \cdot \sigma(\omega),}{\varepsilon_0 \omega (N\cdot d)}=\varepsilon^{1L}_{\text{gr}}(\omega),
\end{equation}
where $ d \approx 0.33\,\rm{nm} $ is the effective thickness of a monolayer and $ \varepsilon_{\text{eff}} $ is the average permittivity of the surrounding medium.

\begin{figure}[ht!]
    \centering
    \includegraphics[width=0.9\linewidth]{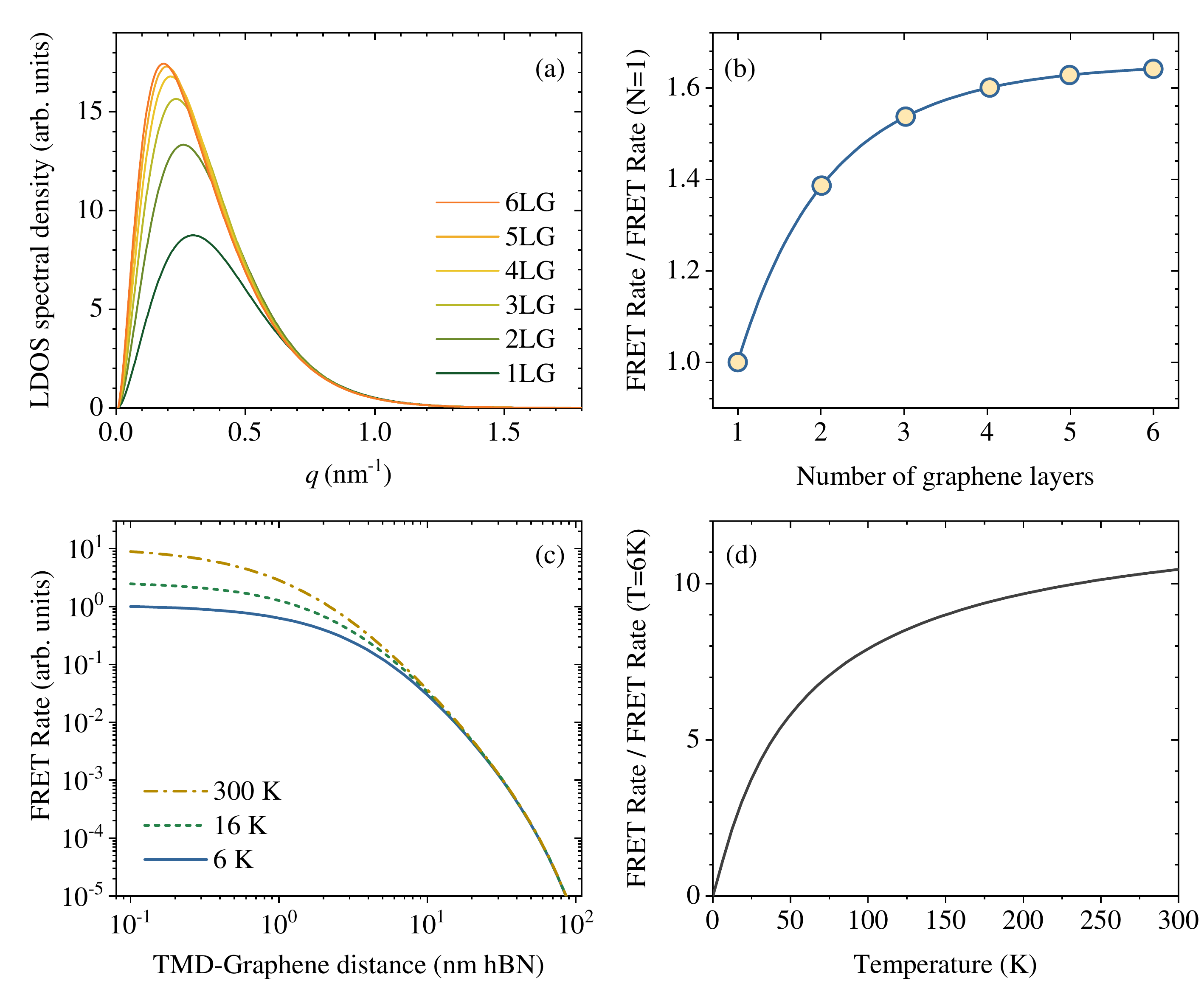}
    \caption{
    (a) Normalized LDOS spectral density as a function of in-plane wavevector $q$ for a dipole embedded in monolayer MoSe$_2$ positioned above a multilayer graphene structure. Different curves correspond to increasing numbers of graphene layers. (b) Total decay rate enhancement (FRET rate) as a function of the number of graphene layer number $ N $. This calculation considers a full integration of the LDOS spectrum in (a), i.e., it considers transfer of hot excitons with finite momentum. This curve is also plotted in Fig.~2 of the main text. (c) FRET rate as a function of the TMD-Graphene distance (in nm hBN) computed at three dfferent temperatures. The TRPL measurements were performed at 6~K. (d) Temperature-dependent FRET rate (normalized the the computed value at 6~K considering a thermalized distribution of $\rm X^0$ excitons in the absence of any spacer between the TMD layer and graphene.  
}
    \label{fig3}
\end{figure}

Using this model, we compute the angular spectrum of the dissipated power,
\begin{equation}
\frac{\qpar}{\mathcal{P}_0} \cdot \frac{\mathrm{d}\mathcal{P}}{\mathrm{d}\qpar}  
\end{equation}

for structures with $ N = 0, 1, \dots, 6 $ graphene layers. The angular spectra (Fig.~\ref{fig3}) reveal how the introduction of graphene (from $N= 0$ to $N=1$ ) vastly increases the high-$ \qpar $ components of the LDOS, which corresponds to near-field energy transfer. We also observe that with increasing number of layers the LDOS does not present any substantial difference.  

To quantify this effect, we compute the total dissipated power by integrating over all in-plane wavevectors:

\begin{equation}
\mathcal{P}^{(N)} = \int_0^\infty \frac{\mathrm{d} \mathcal{P}^{(N)}}{\mathrm{d}\qpar} \cdot \qpar \, \mathrm{d}\qpar.
\end{equation}
The LDOS enhancement is then defined as
\begin{equation}
\frac{\rho(N)}{\rho(0)} = \frac{\mathcal{P}(N)}{\mathcal{P}(0)},
\end{equation}
allowing us to isolate the influence of graphene from other geometric and material effects in the heterostructure.

In principle, the total decay rate  $\Gamma$  of a dipole emitter is proportional to the local density of optical states (LDOS), such that $ \Gamma / \Gamma_0 = \rho / \rho_0 $, where $ \Gamma_0 $ and $ \rho_0 $ refer to the vacuum values. However, this direct proportionality holds only under the assumption that the emitter has a unit quantum yield—that is, all decay proceeds via radiative channels that couple to the electromagnetic environment.

The calculated normalized LDOS angular spectrum is presented in Fig.~\ref{fig3}a for the heterostructure shown in Fig.~\ref{fig2}b. We observe a strong suppression of the FRET contribution as the in-plane wavevector approaches the light cone ($\qpar \to k_0$), as well as variations in the LDOS amplitude within the high-$\qpar$ (FRET-dominated) region as a function of graphene layer number. By integrating the angular spectrum outside the light cone ($\qpar > k_0$), we extract the total FRET-related LDOS, which is plotted as a function of graphene layer number in Fig.~\ref{fig3}d.

To account for the fact that excitons occupy a thermal distribution in momentum space, we introduce an effective LDOS accessible to a thermalized reservoir of direct excitons, denoted as $\mathrm{LDOS}_{\mathrm{eff}}^{X^0}$. We assume that the exciton center-of-mass momentum follows a Boltzmann distribution centered at $\qpar = 0$, which restricts the range of $\qpar$ values contributing to the energy transfer. The occupation probability is given by

\begin{equation}
W(\qpar) = \exp\left[-\frac{E(\qpar)}{k_B T}\right],
\end{equation}

where the exciton dispersion is approximated as

\begin{equation}
E(\qpar) = \frac{\hbar^2}{2M}\left(\qpar - \qpar_{\mathrm{center}}\right)^2,
\end{equation}

with $M$ the exciton center-of-mass mass and $\qpar_{\mathrm{center}} = 0$ for direct excitons. The effective LDOS is then obtained by weighting the angular spectrum with this thermal distribution:

\begin{equation}
\mathrm{LDOS}_{\mathrm{eff}}^{X^0} = \int_0^\infty \frac{\mathrm{d}\mathcal{P}}{\mathrm{d}\qpar} \cdot W(\qpar) \cdot \mathrm{d}\qpar.
\end{equation}

The resulting distance- and temperature dependence of the FRET rate are shown in Fig.~\ref{fig3}c and \ref{fig3}d, respectively. We observe a pronounced reduction of the FRET rate with decreasing temperature, which arises from the increasing momentum mismatch between the thermally populated exciton states (localized near $\qpar \approx 0$) and the high-$\qpar$ modes that dominate the FRET-related LDOS.

\clearpage

\section{\textbf{Rate equation modelling and additional discussion}}
\label{SecRate}

\begin{figure}[ht!]
    \centering
    \includegraphics[width=0.3\linewidth]{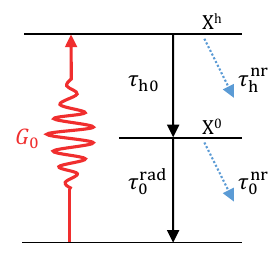}
    \caption{
    Sketch of the three level system discussed in the main text.
}

    \label{Fig_3LS}
\end{figure}

Figure~\ref{Fig_3LS} shows the three level system used in the text to discuss exciton dynamics.
Our model comprises a ground state, a reservoir of finite momentum hot excitons ($\rm X^{\rm h}$), cold, optically active excitons with near-zero momentum ($\rm X^0$). $\rm X^{\rm h}$ are pumped non-resonantly (through phonon assisted processes) at a rate $G_0$ in the continuous wave excitation regime or using femtosecond pulses (for TRPL measurements). $\rm {X^h}$ may relax, with a characteristic time $\tau_{\rm{ho}}$ into $\rm X^0$, which in turn can radiatively recombine (with decay time $\tau_0^{\rm{rad}}$). Non-radiative losses are considered for each population (with decay times $\tau_{\mathrm {h}}^{\mathrm{nr}}$ and $\tau_{\rm {0}}^{\mathrm{nr}}$; respectively). The presence of graphene layer opens non-radiative transfer pathways that may considerably shorten  $\tau_{\rm {h}}^{\mathrm{nr}}$ and $\tau_{\rm {0}}^{\mathrm{nr}}$ and lead to faster PL dynamics and PL quenching. In the following, for each characteristic time $\tau_i$, we associate a decay rate $\Gamma_i=\tau_i^{-1}$. The $\rm {X^0}$ PL intensity is $I_{\rm{X^0}}=\Gamma_0^{\rm{rad}}\:n_{\rm{X^0}}$, where $n_{\rm{X^0}}$ is the $\rm X^0$ population.

~

The rate equations associated with $n_{\rm{X^0}}$ and $n_{\rm{X^h}}$, the hot exciton population are:

\begin{align}
    \dot{n}_{\rm{X^h}}=G_0-\Gamma_{\rm{h}}\:n_{\rm{X^h}}\\
    \dot{n}_{\rm{X^0}}=\Gamma_{\rm{h0}}\:n_{\rm{X^h}}-\Gamma_0\:n_{\rm{X^0}},
\end{align}

With $\Gamma_{\rm{h}}=\Gamma_{\rm{h0}}+\Gamma_{\rm{h}}^{\rm{nr}}$ and $\Gamma_{0}=\Gamma_{\rm{0}}^{\rm{rad}}+\Gamma_{\rm{0}}^{\rm{nr}}$, the total hot and cold exciton population decay rate, respectively.

\textbf{Pulsed regime -- } Under femtosecond pulsed excitation, we get:
\begin{equation}
    n_{\rm{X^0}}\left(t\right)=n_0\frac{\Gamma_{\rm{h0}}}{\Gamma_{\rm{h}}-\Gamma_{0}}\left[ e^{-\Gamma_0\:t}-e^{-\Gamma_{\rm h}\:t}\right].
\label{EqTRPL}
\end{equation}

Eq.\eqref{EqTRPL} is used throughout the manuscript to fit the TRPL data and extract the PL rise and decay time. We note that if $\Gamma_0>\Gamma_{\rm h}$, the rise time will be the $\rm X^0$ lifetime $\tau_0$ and the decay time will be the relaxation time $\tau_{\rm h}$ and vice versa if $\Gamma_0<\Gamma_{\rm h}$~\cite{Fang2019}.

\textbf{Steady state -- } In the steady state we obtain:

\begin{align}
    n_{\rm{X^h}}=\frac{G_0}{\Gamma_{\rm h}}\\
    n_{\rm{X^0}}=\frac{G_0}{\Gamma_0} \times \frac{\Gamma_{\rm{h0}}}{\Gamma_{\rm h}},\\
\end{align}

and thus we get
\begin{equation}
    I_{\rm X^0}=G_0\: \eta_{\rm F}\: \eta_{0},
    \label{EqIX0}
\end{equation}
with $\eta_{\rm F}=\frac{\Gamma_{\rm{h0}}}{\Gamma_{\rm h}}$, the $\rm X^0$ formation yield, $\eta_{0}=\frac{\Gamma_{0}^{\rm{rad}}}{\Gamma_{0}}$, the  $\rm X^0$ emission yield. Eq.\eqref{EqIX0} can then be used to calculate the quenching factor $Q_{\rm{X^0}}$ discussed in the main text. Noteworthy, to evaluate the quenching factor, one needs to assess the ratio of the emission yields 

\begin{equation}
    \frac{\eta_0\left(0\right)}{\eta_0\left(N\right)}=\frac{\Gamma_0^{\rm {rad}}  \left(0\right)}{\Gamma_0^{\rm {rad}}\left(N\right)}\times\frac{\Gamma_0\left(N\right)}{\Gamma_0\left(0\right)},
    \label{Eqkappa0}
\end{equation} 
where $(N)$ refers to MoSe$_2/N$LG  and $(0)$ refers to bare MoSe$_2$.
The second term in Eq.\eqref{Eqkappa0} is directly deduced from the $\rm X^0$ lifetimes measured through TRPL. The first term can be evaluated as in Ref.~\cite{Lorchat2020}, by considering the reduction in exciton binding energy in MoSe$_2/N$LG through hot photoluminescence measurements of excited excitonic states (not shown here) and the fact that the radiative lifetime is proportional to the inverse square of the exciton binding energy. We have observed the $\rm X^0$ radiative lifetime in MoSe$_2/N$LG is approximately 2 times larger than in the MoSe$_2$ reference and that this increase barely depends on $N$, such that 
\begin{equation}
    \frac{\eta_0\left(0\right)}{\eta_0\left(N\right)}\approx 2\times \frac{\tau_{\rm X^0}(0)}{\tau_{\rm X^0}(N)}.
    \label{Eqkappa02}
\end{equation}

Eq.\eqref{Eqkappa02} is then used to compute $\eta(N)/\eta(0)$ from the TRPL measurements in Fig.~2 of the main text. Note that in the main manuscript, for the sake of clarity, the $\rm X^0$ and $\rm{X^h}$ lifetimes are denoted $\tau_{\rm X^0}$ and $\tau_{\rm {X^h}}$, respectively, instead of $\tau_0$ and $\tau_{\rm h}$.

\textbf{Application to hot exciton transfer --} Finally we discuss the details of hot exciton transfer in MoSe$_2/N$LG. We assume that 
\begin{equation}
    \Gamma_{\rm h}\left(N\right)=\Gamma_{\rm{h0}}+\Gamma_{\rm{h}}^{\rm{nr,0}}+\Gamma_{\rm{h}}^{\rm{FRET}}\left(N\right),
\end{equation}

where hot exciton relaxation occurs through $\rm X^0$ formation ($\Gamma_{\rm{h0}}$), non-radiative losses including charge tunneling-mediated phenomena, as well as formation of momentum-dark exciton and other optically inactive excitonic species ($\Gamma_{\rm{h}}^{\rm{nr,0}}$), and F\"orster type resonant energy transfer ($\Gamma_{\rm{h}}^{\rm{FRET}}\left(N\right)$) evaluated in Sec.\ref{SecSotos}. For simplicity, the first two terms are assumed to be independent of $N$, in keeping with the very weak dependence of $\rm X^0$ dynamics on $N$ revealed in Fig.~2. To confront the measurement of the PL intensities in $N$LG with the prediction of our FRET model (Fig.~\ref{fig3}), we consider the quantity:
\begin{equation}
    \frac{Q_{\rm{X^0}}\left(N\right)}{Q_{\rm{X^0}}\left(1\right)}=\frac{I_{\rm X^0}\left(1\right)}{I_{\rm X^0}\left(N\right)}=\frac{\eta_0\left(1\right)}{\eta_0\left(N\right)}\times \frac{\alpha+\beta\left(N\right)}{\alpha+1},
\end{equation}
where $\alpha=\frac{\Gamma_{\rm{h0}}+\Gamma_{\rm{h}}^{\rm{nr,0}}}{\Gamma_{\rm{h}}^{\rm{FRET}}\left(1\right)}$, $\beta\left(N\right)=\frac{\Gamma_{\rm{h}}^{\rm{FRET}}\left(N\right)}{\Gamma_{\rm{h}}^{\rm{FRET}}\left(1\right)}$ and where we have assumed that the absorption rate is independent on $N$.
In the case $\alpha \gg 1$, we obtain $\frac{Q_{\rm{X^0}}\left(N\right)}{Q_{\rm{X^0}}\left(1\right)}\approx \frac{\eta\left(1\right)}{\eta\left(N\right)}$, which remains close to unity based on the results in Fig. 2 of the main manuscript. In contrast, $\alpha \ll 1$, we obtain $\frac{Q_{\rm{X^0}}\left(N\right)}{Q_{\rm{X^0}}\left(1\right)}\approx \beta\left(N\right)$ and we would expect that  $\frac{Q_{\rm{X^0}}\left(N\right)}{Q_{\rm{X^0}}\left(1\right)}$ follows the theoretical predictions in Fig.~\ref{fig3}. In Fig. 2f, the qualitative agreement between our model and experimental results for $\rm Q_{\rm X^0}$ and $\rm Q_{\rm {tot}}$ suggests that FRET may affect significantly hot exciton dynamics.


\center\rule{8cm}{1pt}

\end{document}